\documentclass[aps,prc,reprint,showpacs,groupedaddress,onecolumn]{revtex4}
\usepackage{bm}% bold math
\usepackage{amsmath}
\usepackage{amsfonts}
\usepackage{amscd} 
\usepackage{graphicx}
\usepackage{color}
\def\beq{\begin{equation}}
\def\eeq{\end{equation}}

\begin{document}

\title{Multiresolution decomposition of quantum field theories using wavelet bases}

\author{Tracie Michlin}
\affiliation{Department of Applied Mathematics and Computational Science, 
The University of
Iowa, Iowa City, IA 52242
}

\author{W.~N.~Polyzou}
\affiliation{Department of Physics and Astronomy, The University of
Iowa, Iowa City, IA 52242}

\author{Fatih Bulut}
\affiliation{Department of Physics, In\"on\"u University, 
Malatya, Turkey} 

\date{\today}


\begin{abstract}

  We investigate both the theoretical and computational aspects of using
  wavelet bases to perform an exact decomposition of a local field
  theory by spatial resolution.  The decomposition admits natural
  volume and resolution truncations.  We demonstrate that flow
  equation methods can be used to eliminate short-distance degrees of
  freedom in truncated theories.  The method is tested on a free
  scalar field in one dimension, where the spatial derivatives couple
  the degrees of freedom on different scales, although the method is
  applicable to more complex field theories.  The flow equation method 
  is shown to decouple both distance and energy scales in this example.
  The response to changing the volume and resolution cutoffs and the mass
  is discussed.
\end{abstract}

\maketitle

\section{Introduction}

In this paper we investigate the use of wavelet methods
\cite{best:1994}\cite{federbush:1995}\cite{halliday}\cite{Battle:1999}\cite{best:2000}\cite{Ismail1:2003}\cite{Ismail2:2003}\cite{altaisky:2007}\cite{albeverio:2009}\cite{altaisky:2010}\cite{fatih2}\cite{altaisky:2013}\cite{altaisky:2013b}\cite{fatih1}
\cite{Brennen}\cite{altaisky:2016}\cite{altaisky:2016b}\cite{altaisky:2016c}\cite{Evenbly}\cite{singh}
to decouple degrees of freedom on different distance scales in local
quantum field theory.  Daubechies wavelets and scaling functions can
be used to construct an orthonormal basis of compactly supported
functions
\cite{daubechies:1988}\cite{daubechies}\cite{latto1}\cite{latto2}
\cite{beylkin1}\cite{beylkin2}\cite{beylkin-92}\cite{beylkin3}\cite{kaiser}\cite{resnikoff}\cite{jorgensen}\cite{wavelets}\cite{kessler03}\cite{kessler04}\cite{fatih3}.
The basis includes functions that vanish outside of any arbitrarily
small open set.  They are generated from a single function using
translations and unitary scale transformations.  These basis functions
decompose the Hilbert space into a direct sum of orthogonal subspaces
associated with different resolutions.  Expanding local fields in this
basis leads to an exact representation of the field as an infinite
linear combination of operators with different spatial resolutions.
This expansion replaces the operator-valued distributions by infinite
linear combinations of basis functions with operator-valued
coefficients.  The operator valued coefficients are defined by
smearing the local fields with the basis functions.  While the full
expansion is exact, there are natural volume and resolution
truncations that are defined by retaining only terms in the expansion
that have support intersecting a given volume and with a specified
finest resolution.

We limit our considerations to the use of Hamiltonian
methods, however the representations used in this paper could also be
employed in any field theory framework to provide natural volume and
resolution truncations.   They could also be utilized in alternative
wavelet approaches \cite{altaisky:2007}\cite{altaisky:2010}
\cite{altaisky:2013}\cite{altaisky:2013b}\cite{altaisky:2016}
\cite{altaisky:2016b}
\cite{altaisky:2016c}. When the fields are
replaced by these expansions in field-theory Hamiltonians, local
products of field operators are replaced infinite linear combinations
of products of well-defined operators.  The singularities that arise
from the local operator products reappear as non-convergence of sums,
so the renormalization problem takes on a different form.  The theory
is naturally regularized by truncating the basis in both resolution
and volume.

The problem of constructing a local limit involves first solving the
field equations for truncated theories with different volume and
resolution cutoffs and adjusting the dimensionless parameters of each
truncated field theory to preserve some common observables.  Since the
truncated theories are systems with a finite number of degrees of
freedom, they can in principle be solved, just like lattice
truncations.  The problem is to identify a sequence of truncated
theories and a limiting procedure that results in a well-defined
infinite-volume, infinite-resolution limit that satisfies the axioms
of a local field theory.  The general existence of such a limit is an
unsolved problem, and is beyond the scope of this paper.
%For the case
%of a free scalar field it has been shown that infinite resolution limit of
%truncated correlation functions converge to the Wightman functions of
%the continuum theory \cite{singh}.

However, for measurements involving a fixed energy scale and finite
volume, the number of relevant degrees of freedom is finite.  Under
these conditions both the accessible volume and resolution are
limited.  Truncated field theories that include degrees of freedom
associated with this volume and resolution should describe physics on
this scale after determining the parameters of the truncated theory by
experiment.  The predictions at this scale should be improvable as the
volume and resolution are increased by finite amounts.  This is
independent of the existence of an infinite volume - infinite
resolution limit that describes physical phenomena on all scales.

While it is possible to work at successively finer resolutions, there
are reasons to eliminate short-distance degrees of freedom that are
much smaller than the scales accessible to a given experiment. This
reduces the number of degrees of freedom, where the effects of the
eliminated degrees of freedom appear in a more complicated effective
Hamiltonian that only involves the physically relevant degrees of
freedom.  This is similar in spirit to the program initiated by
Gl\"ockle and M\"uller to eliminate explicit pion degrees of freedom
in a field theory of interacting pions and nucleons \cite{glockle}
using an ``Okobu'' transformation\cite{okobu}.

A feature of the wavelet representation is that the commutation
relations among the field operators are all discrete, and there are
irreducible canonical pairs of operators associated with each
resolution and volume.  The truncated Hamiltonians with different
resolutions have the same form, with coefficients that are rescaled as
a function of the resolution.  There is a natural transformation that
transforms the high-resolution truncated Hamiltonian to the sum of the
corresponding low-resolution truncated Hamiltonian and corrections
that involve the missing high-resolution degrees of freedom.  These
corrections include operators that couple the high and low-resolution
degrees of freedom.
 
Block diagonalizing this Hamiltonian according to resolution gives an
effective Hamiltonian entirely in the low-resolution degrees of
freedom that includes the physics of the eliminated high-resolution
degrees of freedom.  This can be compared to the original
low-resolution Hamiltonian to see how it must be modified to include
the effects of the eliminated degrees of freedom.  In this
representation explicit high-resolution degrees of freedom are
replaced by more complicated effective interactions in the
low-resolution degrees of freedom.  While this process generates new
effective operators, the coefficients of these operators are
well-defined functions of the parameters of the original theory, so in
a renormalizable theory there is no need to introduce new parameters
associated with the new effective operators, although this can always
be done to improve convergence.

In this paper we investigate the use of flow equation methods
\cite{wegner}\cite{glazek93}\cite{glazek2}\cite{perryf}\cite{bartlett}\cite{kehrein}\cite{bogner1}\cite{perry}\cite{bogner2}
to perform the block diagonalization of the high-resolution
Hamiltonian.  In general the flow equation will generate an infinite
collection of complicated effective operators.  In order to separate
the problem of convergence of the flow equation from an analysis of
the scaling properties of the effective interactions, we consider the
case of a free field.  For free fields the different resolution
degrees of freedom are coupled by spatial derivatives, but the
structure of the operators generated by the flow equation remain
quadratic functions of the fields, which restricts the structure of
the operators that are generated by the flow equation to a finite
number of classes.  Because of this, the flow equation can be solved
without addressing the problem of how to manage the generated
effective interactions.  This provides a first test of the proposed
flow equation method to separate scales.

\section{Background - wavelet basis}   

In this section the basis of functions that will be used to expand the
field operators are defined.  The basis functions on the real line are
Daubechies scaling functions
\cite{daubechies:1988}\cite{daubechies}\cite{resnikoff}\cite{wavelets}\cite{kessler03}\cite{kessler04}
on a fixed scale and Daubechies wavelets on all smaller scales.

Our preference for the Daubechies basis is because the basis functions
are orthonormal and have compact support.  The scale is associated
with the size of the support of different basis functions.  In higher
dimensions the basis functions are products of the one-dimensional
basis functions defined in this section.  This leads to a
representation of the theory in terms of local observables.  The
structure of the truncated theory is similar to lattice truncations
which are also formulated in terms of local degrees of freedom.  One
advantage of wavelet truncations is that it is possible to include
independent degrees on different scales, so large-scale degrees of
freedom do not have to be generated by the collective dynamics of many
small-scale degrees of freedom.

One useful property of the scaling-wavelet basis is that all of the
basis functions can be constructed from a single function, $s(x)$,
called the scaling function, by integer translations and dyadic scale
transformations.  The scaling function, $s(x)$, is the
solution of the following linear renormalization group equation

\beq
s(x) = \underbrace{S
\underbrace{(\sum_{l=0}^{2K-1} h_l T^l s(x))}_{\mbox{block average}}
}_{\mbox{rescale}} .
\label{b.1}
\eeq
%\beq
%s(x) = \sum_{l=0}^{2K-1} h_l D T^l s(x) . 
%\label{b.1}
%\eeq
The normalization of the solution of this homogeneous equation is fixed by
the condition
\beq
%\[
\int dx s(x) =1.
%\]
\label{b.2}
\eeq
In equation (\ref{b.1}) $T$ is a unitary integer translation operator
and $S$ is a unitary scale transformation operator that shrinks the
support of a function by a factor of two.  These
operators are
\beq
Ts(x) = s(x-1) \qquad Ss(x) = \sqrt{2} s(2x) .
\label{b.3}
\eeq
Equation (\ref{b.1}) implies that $s(x)$ is the fixed point of the
operation of taking a weighted average of a finite number of
translated of copies of $s(x)$ scaled to half of the original support. 
$K$ is a fixed integer that is related to the smoothness of the basis 
functions.  The
weights, $h_l$, are real numbers determined by the three conditions:
\begin{itemize}
\item [1.] Orthonormality of integer translates of $s(x)$:
\beq
\int s(x) s(x-n) dx = \delta_{n0}.
\label{b.4}
\eeq
\item [2.] Consistency:
\beq
\sum_{l=0}^{2K-1} h_l =\sqrt{2}. 
\label{b.5}
\eeq
\item[3.] Ability to locally pointwise represent low-degree
polynomials:
\beq 
x^n = \sum_{m=--\infty}^{\infty}  c_m s(x-m) \qquad 0 \leq n\leq K.
\label{b.6}
\eeq
\item[] Although the sum in (\ref{b.6}) is infinite, there are no
convergence problems because only a finite number of terms in this
sum are non-zero at any given point.
\end{itemize}

There are two solution of equations (\ref{b.4}-\ref{b.6}) for the
$h_l$. They are related by $h_l'= h_{2K-1-l}$.  The corresponding 
fixed points, $s(x)$, are mirror images of each other.
The resulting $s(x)$ has compact support on the finite interval
$[0,2K-1]$.
The values for 
$K=3$, which are used in this work, are given in table I.  
These $h_l$ are simple algebraic numbers. 

\begin{table}[t]
\caption{Scaling Coefficients for Daubechies K=3 Wavelets}
\begin{tabular}{|l|l|}
\hline				      		      
$h_0$ & $(1+\sqrt{10}+\sqrt{5+2\sqrt{10}}\,)/16\sqrt{2}$ \\
$h_1$ & $(5+\sqrt{10}+3\sqrt{5+2\sqrt{10}}\,)/16\sqrt{2}$ \\
$h_2$ & $(10-2\sqrt{10}+2\sqrt{5+2\sqrt{10}}\,)/16\sqrt{2}$ \\
$h_3$ & $ (10-2\sqrt{10}-2\sqrt{5+2\sqrt{10}}\,)/16\sqrt{2} $ \\
$h_4$ & $(5+\sqrt{10}-3\sqrt{5+2\sqrt{10}}\,)/16\sqrt{2}$ \\
$h_5$ & $(1+\sqrt{10}-\sqrt{5+2\sqrt{10}}\,)/16\sqrt{2}$ \\
\hline
\end{tabular}
\label{coef}
\end{table}

Scaling functions are defined by translating and rescaling $s(x)$:
\beq
%\[
s^k_n (x) := S^k T^n s(x) = 2^{k/2} s\left (2^k(x-2^{-k}n)\right ) .
%\]
\label{b.7}
\eeq
It follows from (\ref{b.4}) and the unitarity of $S$ that 
the functions $s^k_n (x)$ are orthonormal for each fixed $k$.

Subspaces ${\cal S}_k (\mathbb{R}) \subset L^2(\mathbb{R})$
of resolution $1/2^k$ are defined by
%\beq
\[
{\cal S}_k := \{ f(x) \vert f(x)= \sum_{n=-\infty}^\infty c_n s^k_n(x),
\quad \sum_{n=-\infty}^\infty \vert c_n \vert^2 < \infty \}.
\]
%\label{b.8}
%\eeq
It follows from (\ref{b.1}) that these subspaces are related by 
%\beq
\[
{\cal S}_k := S^k {\cal S}_0
\qquad
{\cal S}_{k} \subset {\cal S}_{k+n} \qquad n\geq 0 
\]
%\label{b.9}
%\eeq
or more generally they are nested
\beq
\cdots {\cal S}_{k-1} \subset {\cal S}_{k} \subset {\cal S}_{k+1} \subset \cdots .
\label{b.10}
\eeq
The inclusions in (\ref{b.10}) are proper in the sense that they
have non-empty orthogonal complements
%\beq
\[
{\cal S}_{k+1} = {\cal S}_{k} \oplus {\cal W}_{k} \qquad {\cal W}_k \not= \{\emptyset \}.   
\]
%\label{b.11}
%\eeq
The space 
${\cal W}_{k}$ is the orthogonal complement of ${\cal S}_{k}$
in ${\cal S}_{k+1}$.  From a physical point of view
${\cal S}_{k+1}$ is a finer resolution subspace than
${\cal S}_{k}$, and ${\cal W}_{k}$ fills in the missing degrees of
freedom that are in ${\cal S}_{k+1}$ but not in ${\cal S}_{k}$.
Combining these decompositions we have the following relation
between the subspaces ${\cal S}_{k+n}$ and ${\cal S}_{k}$ of different
resolutions
\beq
{\cal S}_{k+n} = {\cal S}_{k}\oplus {\cal W}_{k} \oplus {\cal W}_{k+1}
\oplus \cdots \oplus {\cal W}_{k+n-1} .
\label{b.12}
\eeq
%The subpaces ${\cal W}_{k}$ represents the fine-resolution 
%details that fall between the subspaces ${\cal S}_{k+1}$ and ${\cal S}_{k}$
The limit of this chain as $n \to \infty$ leads to an exact 
decomposition of $L^2 (\mathbb{R})$ by resolution:  
\beq
L^2 (\mathbb{R}) = {\cal S}_{k}\oplus {\cal W}_{k} \oplus {\cal W}_{k+1}
\oplus {\cal W}_{k+2} \oplus {\cal W}_{k+3} \oplus  \cdots .
\label{b.13}
\eeq
The subspaces ${\cal W}_{k}$ are called wavelet spaces.  Orthonormal bases
for the subspaces ${\cal W}_{k}$ are constructed from the mother wavelet,
$w(x)$, which is defined by taking a different weighted average of
translations of the scaling function $s(x)$ scaled to half of the support
of $s(x)$:
\beq
w(x) := \sum_{l=0}^{2K-1} g_l  S T^l s(x)
\qquad
g_l = (-)^l h_{2K-1-l} .
\label{b.14}
\eeq
The weights $g_l$ in equation (\ref{b.14}) are related to the weights $h_l$
used in (\ref{b.1}) except the signs alternate and the order of
the indices is reversed.

Applying powers of the dyadic scale transformation
operator, $S$,  and integer translation operator, $T$, to 
$w(x)$ gives the following basis functions for ${\cal W}_k:$
\beq
w^k_m(x) := S^k T^m w(x) = 2^{k/2} w \left (2^k (x-2^{-k}m)\right ) .
\label{b.15}
\eeq
The functions $w^k_m(x)$ are called wavelets. 
It can be shown that for each fixed $k$,
$\{ w^k_m(x) \}_{m=-\infty}^\infty$ is an
orthonormal basis for the subspace ${\cal W}_{k}$.   Because of
(\ref{b.12}) the $w^k_n(x)$ for different values of $k$ are also
orthogonal.

From (\ref{b.7}),(\ref{b.14}) and (\ref{b.15}) it follows that both $s^k_m(x)$ and
$w^k_m(x)$ can be constructed from the fixed point, $s(x)$, of the
renormalization group equation (\ref{b.1}), using elementary
transformations.

The decomposition (\ref{b.13}) implies that for any
fixed starting scale $2^{-k}$,
%\beq
\[
\{ s^k_n(x) \}_{n=-\infty}^{\infty} \cup 
\{ w^l_n(x) \}_{n=-\infty, l=k}^{\infty,\infty}.
\]
%\label{b.16}
%\eeq
is an orthonormal basis for $L^2(\mathbb{R})$ consisting of compactly
supported functions.  The support of both $s^k_m(x)$ and $w^k_m(x)$ is
$[2^{-k} m , 2^{-k}(m+ 2K-1)]$.  For any point on the real line there
are basis functions of arbitrarily small support that include that
point.

The basis functions $s^k_n(x)$ are associated with degrees of freedom
of scale $2^{-k}$ and the $w^l_n(x)$ are associated with degrees of
freedom of scale $2^{-(l+1)}$ that are not of scale $2^{-l}$.  Thus
they are identified with localized degrees of freedom with distance
scales $2^{-k-l}$ for all integers, $l\geq 0$.

Equation (\ref{b.12}) implies that the 
functions
%\beq
\[
\{ s^{k+m}_n (x)\}_{n=-\infty}^\infty
\qquad \mbox{and} \qquad 
\{ s^k_n(x)\}_{n=-\infty}^{\infty} \cup 
\{ w^l_n(x) \}_{n=-\infty, l=k}^{\infty,k+m-1}
\]
%\label{b.17}
%\eeq
are related by an orthogonal transformation.  This transformation is
called the wavelet transform.  It can be computed more efficiently
than a fast Fourier transform, using the $h_l$ and $g_l$ as weights that define
``low''- and ``high-pass'' filters
%\beq
\[
s_n^{k-1}(x) = \sum h_l s^k_{2n+l}(x) 
\qquad 
w_n^{k-1}(x) = \sum g_l s^k_{2n+l}(x) .
\]
%\label{b.18}
%\eeq
The inverse of this orthogonal transformation is  
%\beq
\[
s^{k}_n (x) = \sum_{m} h_{m-2n} s^{k-1}_{m} (x)
+ \sum_m g_{m-2n} w^{k-1}_{m} (x).
\]
%\label{b.19}
%\eeq

It is precisely these transformations (or their three-dimensional
generalization) that relate a fine-resolution Hamiltonian to the sum
of a coarse resolution Hamiltonian plus fine scale corrections.

The Daubechies wavelets and scaling functions are fractal functions.
This is because $s(x)$ is the solution of a renormalization group
equation, and all of the basis functions are obtained by applying a
finite number of scale transformations, translations and sums to
$s(x)$.  In spite of their fractal nature, these basis functions have
a finite number of derivatives that increase with increasing $K$.
This paper uses the $K=3$ basis. These basis functions have one
continuous derivative.  This allows for an exact representation of
Hamiltonians that have fields with at most one derivative.  It
explicitly avoids the need for finite difference approximations to
derivatives.  Increasing $K$ leads to smoother basis functions at the
expense of larger support and increasing overlap with basis functions
on the same scale.   

Quantum fields are generally assumed to be operator-valued
tempered distributions. This suggests that field operators smeared
with test functions that only have a finite number of derivatives
might not be well-defined operators, however free-field Wightman functions
smeared with Daubechies $K\geq 3$ wavelets or scaling functions are
well-defined.  This follows from the analytic expressions for the free
field Wightman functions \cite{bogoliubov} and the fact that the basis
function have compact support and a continuous derivative.  This is
analogous to the observation that a delta function, which is a
distribution, is also a well-defined linear functional on the space of
continuous functions.  Equation (\ref{b.6}) shows that certain linear
combinations of these functions can be much smoother.  In reference
\cite{singh} it is shown specifically that vacuum expectation values
of the Daubechies' wavelet smeared free fields converge to the exact
free-field Wightman functions in the limit of infinite resolution.
For theories truncated to a finite number of degrees of freedom, the
Stone-von Neumann theorem \cite{stone}\cite{vonneumann}, which
establishes the unitary equivalence of all representations of the
canonical commutation relations, makes it possible to formulate the
dynamics of the truncated theory in terms of the well-defined algebra
of wavelet-smeared free fields on the free-field Fock space.  This
ensures that the fractal nature of the basis does not cause any
problems in the treatment of the truncated theory.  If there are any
issues with the fractal nature of the wavelet basis functions, they
must arise when one tries to establish the existence of a local limit.

Since the Daubechies basis functions have compact support,
their Fourier transforms are analytic.  Thus, expanding the field in
a coordinate-space wavelet basis is equivalent to expanding the
Fourier transform of the field in an analytic basis.    
The Fourier transformed basis functions are infinitely differentiable.
They fall off like inverse powers of the momentum, similar to Feynman diagrams.
None of them have compact support.

Another potential issue with fractal basis functions involves their
computation.  This turns out to be a non-issue because they have
compact support and integrals of products of these functions with
polynomials of arbitrarily high-degree can be computed exactly
(reduced to finite linear algebra) using the renormalization group
equation (\ref{b.1}).  Since any continuous function on a compact
interval can be approximated by a polynomial, it is possible to
accurately compute integrals of products of these basis functions with
any continuous function.  The renormalization group equation can also
be used to reduce the computation of arbitrary products of these basis
function and their derivatives to finite linear algebra.  It is even
possible to use these methods to evaluate integrals of products of
these basis functions with functions having logarithmic or
principal-value singularities
\cite{wavelets}\cite{kessler03}\cite{kessler04}\cite{fatih3}.  The
computational methods relevant to this work are discussed in the
appendix.

\section{Wavelet discretized fields}

Given a pair of scalar fields $\Phi (x)$ and $\Pi (x)$
satisfying canonical equal-time commutation relations
%\beq
\[
[\Pi (\mathbf{x},t), \Phi (\mathbf{y},t)]=
-i \delta (\mathbf{x}-\mathbf{y})
\]
%\label{f.1}
%\eeq
%\beq
\[
[\Phi (\mathbf{x},t), \Phi (\mathbf{y},t)]=
[\Pi (\mathbf{x},t), \Pi (\mathbf{y},t)]=0 , 
\]
%\label{f.2}
%\eeq
discrete fields satisfying the discrete form of these 
commutation relations can be constructed by smearing the spatial
coordinates of the fields with an orthonormal set of basis functions.

For the scaling-wavelet basis (we consider the $1+1$  dimensional case
for notational simplicity) the discrete fields are defined by
%\beq
\[
\Phi^k (s,{n},t) := \int d{x} 
\Phi ({x},t) s_{{n}}^{k} ({x})
\qquad
\Phi^l (w,{n},t) := 
\int d{x} 
\Phi ({x},t)
w^l_{{n} } ({x}) \qquad (l \geq k)
\]
%\label{f.3}
%\eeq
%\beq
\[
\Pi^k (s,{n},t) := \int d{x} 
\Pi ({x},t) s_{{n}}^{k} ({x})
\qquad
\Pi^l (w,{n},t) := 
\int d{x} 
\Pi ({x},t)
w^l_{{n} } ({x}) \qquad (l \geq k ) . 
\]
%\label{f.4}
%\eeq
These fields represent degrees of freedom localized on the
support of the associated basis function.

As a result of the orthonormality of the basis functions the 
discrete equal-time commutators are:
\beq
[\Phi^k (s,{n},t),\Phi^k (s,{m},t)]=0
\qquad
[\Pi^k (s,{n},t),\Pi^k (s,{m},t)]=0
\label{f.5}
\eeq
\beq
[\Phi^k (s,{n},t),\Pi^k (s,{m},t)]=i \delta_{{n}{m}}
\label{f.6}
\eeq
\beq
[\Phi^r (w,{n}, t),\Phi^s (w,{m}, t)]=0
\qquad
[\Pi^r (w,{n},t),\Pi^s (w,{m}, t)]=0
\label{f.7}
\eeq
\beq
[\Phi^r (w,{n},t),\Pi^s (w,{m},t)]=i 
\delta_{rs} \delta_{{n}{m}}
\label{f.8}
\eeq
\beq
[\Phi^r (w,{n}, t),\Phi^k (s,{m}, t)]=0
\qquad
[\Pi^r (w,{n}, t),\Pi^k (s,{m}, t)]=0
\label{f.9}
\eeq
\beq
[\Phi^r (w,{n}, t),\Pi^k (s,{m}, t)]=0
\qquad
[\Pi^r (w,{n}, t),\Phi^k (s, {m}, t)]=0 .
\label{f.10}
\eeq
The field operators have the {\it exact} representation
in terms of these discrete operators
\beq
%\[
\Phi ({x},t) =
\sum_{{n}} \Phi^k (s,{n},t) s_{{n}}^{k} ({x}) 
+ \sum_{l \geq k;n} 
\Phi^l (w,{n},t) w^l_{{n} } ({x}) 
%\]
\label{f.11}
\eeq
\beq
%\[
\Pi ({x},t) =
\sum_{{n}} \Pi^k (s,{n},t) s_{{n}}^{k} ({x}) 
+ \sum_{l \geq k ; n} 
\Pi^l (w,{n},t) w^l_{{n}} ({x}). 
%\]
\label{f.12}
\eeq

These expansion can be inserted in the free field Hamiltonian,
\beq
H= 
{1 \over 2}\int ({\Pi}({x},0){\Pi}({x},0)
+\pmb{\nabla}\Phi ({x},0) \cdot \pmb{\nabla}\Phi ({x},0)
+ \mu^2 \Phi({x},0)\Phi({x},0))d{x} ,
\label{f.13}
\eeq
which can be expressed exactly in terms of the $t=0$
discrete fields.  The discrete form of the
exact Hamiltonian is the sum of an operator
with only scaling functions fields, $H_s$, one with only
wavelet fields, $H_w$ and one that has products of
both types of fields, $H_{sw}$:
\beq
%\[
H = H_s + H_w +H_{sw}
%\]
\label{f.14}
\eeq
where
\[
H_s :=  
{1 \over 2} (\sum_{{n}} {\Pi}^k(s,{n},0){\Pi}^k(s,n,0)
+\sum_{{m}{n}} \Phi^k (s,{m} ,0) D^k_{s;{m}{n}} \Phi^k (s,{n},0)
\]
\beq
%\[
+ \mu^2 \sum_{{n}} {\Phi}^k(s,{n},0){\Phi}^k(s,{n},0)),
%\]
\label{f.15}
\eeq
\[
H_w :=  
{1 \over 2} (\sum_{{n},l} {\Pi}^{l}(w,{n},0)
{\Pi}^l(w,{n},0)
+ \sum_{{m},l,{n},j}\Phi^l (w,{m} ,0) 
D^{lj}_{w;{m}{n}}
\Phi^j (w,{n},0)
\]
%\beq
\[
+ \mu^2 \sum_{l,{n}} {\Phi}^l(w,{n},0)
{\Phi}^l(w,{n},0)),
\]
%\label{f.16}
%\eeq
%\beq
\[
H_{sw} :=  
{1 \over 2} 
\sum_{{m},l,{n}}\Phi^l (w,{m} ,0) 
D^{lk}_{sw;{m}{n}}
\Phi^k (s,{n},0) .
\]
%\label{f.17}
%\eeq
The coefficients
$D^k_{s{m}{n}}$,
$D^{lj}_{w,{m},{n}}$ and 
$D^{lk}_{sw;{m},{n}}$ 
that couple near neighbor fields and fields with different scales
are the constant matrices given by
\beq
D^k_{s;{m}{n}}  = \int d {x} 
{d \over dx} s^k_{{m}} ({x})  
{d \over dx} s^k_{{n}} ({x})
\label{f.18}
\eeq
\beq
D^{lj}_{w;{m}{n}}= 
\int d {x} 
{d \over dx} w^l_{{m}} ({x})  
{d \over dx} w^j_{{n}} ({x})
\label{f.19}
\eeq
\beq
D^{lk}_{ws;{m}{n}}=
2 \int d {x} 
{d \over dx} w^l_{{m}} ({x})  
{d \over dx} s^k_{{n}} ({x}) .
\label{f.20}
\eeq

The support properties of the basis functions imply that the matrices
$D^x_{y;mn}$ vanish if the support of the functions in the integrand 
have empty intersection,  so they have a
structure similar to a finite difference approximation.  For a free
field the matrices $D^{lk}_{ws;mn}$ and $D^{lj}_{w,{m},{n}}$ for
$l\not=j$ are responsible for coupling of physical degrees of freedom
on different resolution scales.  In interacting theories there are
additional couplings that come from local products of more than two
fields.  For example
%\beq
\[
\int \phi^4 (x,t) dx = \sum_{n_1 n_2 n_3 n_4}
\Gamma^k_{s;n_1\cdots n_4} \Phi^k(s,n_1,t) \Phi^k(s,n_2,t)
\Phi^k(s,n_3,t) \Phi^k(s,n_4,t) +\cdots
\]
%\label{f.21}
%\eeq
where
\beq
\Gamma^k_{s;n_1\cdots n_4}:= \int s^k_{n_1} (x) s^k_{n_2} (x) s^k_{n_3}
(x) s^k_{n_4} (x) dx
\label{f.22}
\eeq
and the $\cdots$ represent additional terms in the sum that also involve the
wavelet basis functions and fields.  Like the $D^k_{s;mn}$, the coefficients
$\Gamma^k_{s,n_1\cdots n_4}$ are almost local in the sense that
they vanish unless all of the functions in the integral (\ref{f.22}) have
overlapping support.  They also include operators that couple 
degrees of freedom on different scales.

The other important feature of the fractal nature of the
scaling -wavelet basis is that these constant coefficients
have simple scaling properties.  For example
\beq
%\[
D^{k}_{s;mn} = 2^{2k} D^0_{s;mn} = 2^{2k} D^0_{s;0,n-m}
%\]
\label{f.23}
\eeq
\beq
%\[
\Gamma^k_{s;n_1\cdots n_4} = 2^{k} \Gamma^0_{s;n_1\cdots n_4}=
2^{k} \Gamma^0_{s;0,n_2-n_1,n_3-n_1,n_4-n_1}.
%\]
\label{f.24}
\eeq
where we have used translational invariance to express
these coefficients in terms 
of $D^0_{s;0,n-m}$ and
$\Gamma^0_{s;0,n_2-n_1,n_3-n_1,n_4-n_1}$.  In addition, these constant
coefficients can all be computed exactly (i.e. reduced to finite linear
algebra) using (\ref{b.2}-\ref{b.3}) and the scaling equation (\ref{b.1}).  This
is discussed in detail in \cite{fatih2} and the appendix.   See also
\cite{beylkin-92}\cite{beylkin3}\cite{latto1}\cite{latto2} for general
methods to compute integrals involving wavelets and scaling functions.
The result is that for a free field all of the coupling coefficients
can be expressed in terms of the nine non-zero coefficients
$D^0_{s;0m}$ with $-4 \leq m \leq 4$.  These can be computed exactly 
\cite{beylkin-92}\cite{fatih2}.  The results are rational numbers.
Their computation is discussed in the appendix.

\section{Flow equation}

The wavelet basis decomposes the field into a sum of operators that
are localized in different finite volumes.  Each of these operators
are also associated with different resolutions.  For free fields the
coefficients $D^{kl}_{sw;mn}$, $D^{kl}_{ws;mn}$, and $D^{rs}_{ww;mn}$
in the Hamiltonian couple degrees freedom with different resolutions.

One can think of $1/2^k$ as the physically relevant resolution
scale. The canonical scaling function fields $\Phi^k(s,n,t)$ and
$\Pi^k(s,n,t)$ are an irreducible set of operators for degrees of freedom 
on this scale.
The wavelet fields also appear in the Hamiltonian; they are associated
with finer-scale degrees of freedom.  Finally products of wavelet and
scaling function fields represent terms that couple degrees of freedom
on the physical scale to those on smaller scales.

From a physics point of view, while the smaller scales may not be
experimentally relevant, they may represent important contributions
to the dynamics.  One can imagine integrating them out in a functional
integral representation to get an effective theory involving only the
experimentally relevant degrees of freedom.   This is a difficult
calculation in the wavelet representation.

A more direct approach would be to decouple the scaling function part
of the Hamiltonian from the wavelet part.  This would also lead to an
effective Hamiltonian involving only the physically relevant degrees
of freedom $\Phi^k(s,n,t)$ and $\Pi^k(s,n,t)$ and a complementary
Hamiltonian that acts only on the remaining degrees of freedom.  The
decoupling will necessarily generate more complicated effective
interactions among the physically relevant degrees of freedom.

We also remark that the free field Hamiltonian(\ref{f.13}) is still a
many-body Hamiltonian.  Decoupling at the operator level is a stronger
condition than decoupling on a finite number of particle subspace.

Flow equations were introduced by Wegner \cite{wegner} as a method to
continuously evolve a Hamiltonian to a unitarily equivalent
simpler form. 
Flow equations methods
\cite{glazek93}
\cite{glazek2}
\cite{perryf}
\cite{bartlett}
\cite{kehrein}
\cite{bogner1}
\cite{bogner2}
\cite{perry}
are an alternative to direct diagonalization or block
diagonalization methods \cite{okobu}.  They have been applied to
problems in quantum field theory and quantum mechanics.
They have the advantage that
they are simpler to implement than integrating out short-distance
degrees of freedom in a functional integral.
Flow equations are designed to 
perform this diagonalization using a continuously parameterized
unitary transformation, $U(\lambda)$.  The transformed 
Hamiltonian has the form
%\beq
\[
H(\lambda) = U(\lambda) H U^{\dagger}(\lambda).
\]
%\label{feq.1}
%\eeq
Here $H(0)=H$ is the original Hamiltonian; the generator of the flow
equation is chosen to continuously evolve the initial Hamiltonian into
the desired form as $\lambda$ increases.  Here $\lambda$ is called the
flow parameter.  As $\lambda$ increases from $0$ the Hamiltonian
evolves towards the desired form.  The evolution is constructed to
exponentially approach the desired form, but it is possible for the
exponent to become small.  Nevertheless, evaluating $H(\lambda)$ at
any value of $\lambda$ still yields a Hamiltonian that is unitarily
equivalent to the original Hamiltonian with weaker scale coupling
terms.

The preference for flow equations methods in the wavelet representation
is that the simple form of the commutators of the discrete canonical fields,
(\ref{f.5}-\ref{f.10}), reduces the integration of the flow equation to
simple algebra.  The problem is to find a generator of the flow
that leads to the desired outcome.

In general the unitarity of $U(\lambda)$ implies that it satisfies the
differential equation
%\beq
\[
{dU(\lambda) \over d\lambda} = {dU(\lambda) \over d\lambda}
U^{\dagger}(\lambda) U(\lambda) =
K(\lambda) U(\lambda)     
\]
%\label{feq.2}
%\eeq
where
\beq
%\[
K(\lambda) = {dU(\lambda) \over d\lambda}U^{\dagger}(\lambda) = -
K^{\dagger}(\lambda) 
%\]
\label{feq.3}
\eeq
is the anti-Hermitian generator of this unitary transformation.
We are free to choose a generator that leads to the desired
outcome.
%$K(\lambda)$ is anti-Hermitean because 
%\beq
%0 = {d \over d\lambda} (U(\lambda) U^{\dagger}(\lambda)) 
%= {d U(\lambda)\over d\lambda}  U^{\dagger}(\lambda)+    
%U(\lambda){dU^{\dagger}(\lambda) \over d\lambda} 
%\label{feq.4}
%\eeq
%which leads to 
%\beq
%{d U(\lambda)\over d\lambda} = K(\lambda)U(\lambda)
%\qquad
%{d U^{\dagger} (\lambda)\over d\lambda} = - U^{\dagger} (\lambda) K(\lambda) .
%\label{feq.5}
%\eeq
It follows that $H(\lambda)$ satisfies the differential equation
%\beq
\[
{dH(\lambda) \over d\lambda}= [K(\lambda),H(\lambda)] .
\]
%\label{feq.6}
%\eeq
For this application it is useful to choose a generator, $K(\lambda)$, that
is a function of the evolved Hamiltonian:
%\beq
\beq
K(\lambda) = [ G(\lambda), H(\lambda)]
\label{feq.7}
\eeq
where $G(\lambda)$ is the part of $H(\lambda)$ with the operators
that couple different scales turned off.  With this choice 
$G(\lambda)=G^{\dagger}(\lambda)$ so $K(\lambda)$ is anti-Hermitian.

It follows that
\beq
{dH(\lambda) \over d\lambda}= [K(\lambda),H(\lambda)] = [[G(\lambda),H(\lambda)],H(\lambda)] =
[H(\lambda),[H(\lambda),G(\lambda)]] .   
\label{feq.8}
\eeq
Equation (\ref{feq.8}) is the desired flow equation for our free field
Hamiltonian.  A fixed point, $\lambda^*$, of this equation occurs when 
%\beq
\[
[H(\lambda^*) ,[H(\lambda^*),G(\lambda^*)]] = 0 .
\]
%\label{feq.9}
%\eeq

It follows from the 
structure of the equation that the generator $K(\lambda)$   only contains
operators that couple the wavelet and scaling functions degrees of freedom.
%if $H(0)$ is the free field Hamiltonian (\ref{f.13}).  
Below we 
discuss the argument that this non-linear equation drives this commutator
to zero for the case of a free field Hamiltonian.
%solution to such a fixed point.
%We remark that the parameter $\lambda$ in equation (\ref{feq.8})  
%has dimensions of $(energy)^{-2}$.
%
%If we divide $H$
%by a constant $\eta$ with the dimensions of energy then the 
%flow equation has the form
%
%\beq
%   {d(H(\lambda)/\eta) \over d\lambda}= [K(\lambda),H(\lambda)] =
%   [[G(\lambda),H(\lambda)],H(\lambda)] =
%\eta^2 [H(\lambda)/\eta,[H(\lambda)/\eta,G(\lambda)/\eta]] .   
%\label{feq.10}
%\eeq
%If we let $\rho= \eta^2 \lambda$ then the evolution has the for
%\beq
%{d(H(\rho)/\eta) \over d\rho}= 
%[H(\rho)/\eta,[H(\rho)\eta,G(\rho)\eta]]
%\label{feq.11}
%\eeq
%Once we solve for the dimensionless Hamiltonian we have to multiply the 
%result by $\eta$ to get the Hamiltonian back. 

%It is not automatic that solving the flow equation (\ref{feq.8}) will lead to
%a transformed Hamiltonian with no or reduced coupling terms.
%As a practical matter we use the flow equation to reduce the coupling

The following considerations are limited to the case of a free field.
This is because for interacting fields integrating the flow equation
generates an infinite number of new operators. The
non-zero commutators of polynomials of field operators of degree $n$
and $m$ are polynomials of degree $n+m-2$.  Each iteration of the
flow equation increases the degree of the polynomials.  The
new operators represent many-body interactions in the transformed
Hamiltonian.  A separate analysis of the scaling properties of these
many-body polynomial operators is needed to determine the relative
strength of these operators, and which, if any operators can be safely
discarded.  This analysis is separate from considerations about the
flow equation and needs to be developed in applications to realistic
systems.

To understand what happens in the case of the free field
Hamiltonian (\ref{f.13}) first note that for the starting Hamiltonian
all of the operators are quadratic in the $\Phi^k(s/w,n,t) $ and
$\Pi^k(s/w,n,t)$ operators, and commutators of these quadratic 
polynomials remain quadratic polynomials. 
The commutator of the scaling function or
wavelet part of the Hamiltonian with the scale coupling part is either zero or
replaces one scaling function operator with a wavelet function
operator, or one wavelet function operator with a scaling function
operator, resulting in another scale-coupling operator.  Likewise the
commutator of different scale coupling
operator parts gives zero or a product of two
scaling or two wavelet function operators.  This allows us to
separate the flow
equation into separate equations for the scale coupling term
$H_{sw}(\lambda)$ and uncoupled terms
$G(\lambda)= H_s (\lambda) + H_w(\lambda)$.  To do this
consider the equations
%\beq
\[
H(\lambda) = G(\lambda) + H_{sw} (\lambda),  
\]
%\label{feq.12}
%\eeq
%\beq
\[
[G(\lambda), H_{sw} (\lambda)] =: H_{sw}(\lambda)',
\]
Here the prime indicates a different scale coupling operator.
Commuting this operator with $H(\lambda)$ gives
%\label{feq.13}
%\eeq
%\beq
\[
[G(\lambda) + H_{sw} (\lambda),H_{sw}'(\lambda) ] =: 
H_{sw}''(\lambda) + [H_{sw}(\lambda),H_{sw}'(\lambda)]
\]
%\label{feq.14}
%\eeq
where
%\beq
\[
H_{sw}''(\lambda) = [G (\lambda) ,H_{sw}'(\lambda) ] =
[G (\lambda) ,[G(\lambda), H_{sw} (\lambda)]].
\]
%\label{feq.15}
%\eeq
is another scale coupling operator.
The commutator of the scale-coupling terms gives a sum of scaling 
and wavelet terms 
%\beq
\[
[H_{sw}(\lambda),H_m'(\lambda)] = H_s''(\lambda) + H_w''(\lambda).
\]
%\label{feq.16}
%\eeq
This can be used to identify the terms on the right hand side of the 
scaling equation that couple degrees of freedom on different scales and those
that map degrees of freedom on the same scale into themselves.
Defining 
%\beq
\[
H_A (\lambda) = G(\lambda)
\qquad 
H_B (\lambda) = H_{sw} (\lambda), 
\]
%\label{feq.17}
%\eeq
the flow equations can now be separated into coupled equations 
for the mixed (B) and non-mixed (A) parts of the Hamiltonian
%\beq
\[
{dH_A (\lambda) \over d\lambda} = [H_B(\lambda), [ H_B(\lambda), H_A(\lambda)]],
\]
%\label{feq.18}
%\eeq
%\beq
\[
{dH_B (\lambda) \over d\lambda} = [H_A(\lambda), [ H_B(\lambda),
H_A(\lambda)]] = - [H_A(\lambda), [ H_A(\lambda), H_B(\lambda)]] .
\]
%\label{feq.19}
%\eeq
%\beq
%    {dG (\lambda) \over d\lambda} = [H_{sw}(\lambda), [ H_{sw}(\lambda),
%        G(\lambda)]]
%\label{feq.20}
%\eeq
%\beq
%   {dH_{sw} (\lambda) \over d\lambda} = [G(\lambda), [ H_{sw}(\lambda),
%       G(\lambda)]] = - [G(\lambda), [ G(\lambda), H_{sw}(\lambda)]] .
%\label{feq.21}
%\eeq
These equations have a
symmetric form under $H_A(\lambda) \leftrightarrow H_B(\lambda)$ 
except for a sign, which can be seen by changing the
order in the commutator in the second equation.

To understand how these equations evolve the Hamiltonian to the desired
form, we express the first equation
in a basis of eigenstates of $H_B(\lambda)$ with eigenvalues $e_{bn}(\lambda)$ and the second 
in a basis of eigenstates of $H_A(\lambda)$ with eigenvalues 
$e_{an}(\lambda)$.  The equations for the matrix elements in each of these
bases have the form
\beq
%\[
{dH_{Amn} (\lambda) \over d\lambda} =
(e_{bm}(\lambda)-e_{bn}(\lambda))^2 H_{Amn}(\lambda)
%\]
\label{feq.20}
\eeq
and
\beq
%\[
{dH_{Bmn} (\lambda) \over d\lambda} =
-(e_{am}(\lambda)-e_{an}(\lambda))^2 H_{Bmn}(\lambda) .
%\]
\label{feq.21}
\eeq
These equations can be integrated exactly
\beq
H_{Amn}(\lambda) = e^{\int_0^\lambda (e_{bm}(\lambda')-e_{bn}(\lambda'))^2 d\lambda'}
H_{Amn}(0)
\eeq
\beq
H_{Bmn}(\lambda) = e^{-\int_0^\lambda (e_{am}(\lambda')-e_{an}(\lambda'))^2 d\lambda'}
H_{Bmn}(0).
\eeq
These solutions show that matrix elements of $H_A$ increase
exponentially while matrix elements of $H_B$ decrease exponentially as
the flow parameter $\lambda$ is increased.  This evolution
can stall if there are
degeneracies in the eigenvalues, approximate degeneracies, or if
eigenvalues parameterized by $\lambda$ cross.

It is also apparent from these equations that in the high resolution - large 
volume limit, where the spectrum of the block diagonal operators approaches a
continuous spectrum, there will be closely spaced eigenvalues, which
will lead to slow convergence of some parts of the scale coupling
operator.

To solve these equation numerically the Hamiltonian needs to be
truncated to finite number of degrees of freedom.  This means that it
is necessary to truncate both the volume and resolution.  Any system
with a finite energy in a finite volume is expected to be dominated by
a finite number of degrees of freedom\cite{pperry}.  These can be separated into
degrees of freedom associated with an experimental scale and
additional relevant degrees of freedom at smaller scales.  We can use
scaling function fields as the degrees of freedom on the experimental
scale and wavelet degrees of freedom the smaller scales that are still
relevant to the given volume and energy scale.

While similar remarks apply to Hamiltonians with interactions, in
general a different flow generator may be needed separate the desired
degrees of freedom.  It may also be necessary to first project the
truncated Hamiltonian on a subspace before solving the flow equation.

\section{test}

To determine if solving the flow equation eliminates the coupling
terms, we consider a truncation of the free field Hamiltonian
(\ref{f.13}) to a finite volume with two resolutions, using 32 basis
functions - 16 scaling functions and 16 wavelets to expand the fields.
For simplicity, we only keep wavelets on one scale.
The coefficients $D^k_{s;{m}{n}}$, $D^{lj}_{w;{m}{n}}$ and
$D^{lk}_{sw;{m}{n}}$, in (\ref{f.18}),(\ref{f.19}),(\ref{f.20}) are
given in \cite{fatih2} and computed in the appendix.
The truncated fields are defined by an expansion in a finite number of 
basis functions of two resolutions:
\beq
%\[
\Phi (x) = \sum_{n=0}^{15} s_n(x) \Phi (s,n,t) + \sum_{n=0}^{15} w_n (x)
\Phi (w,n,t)
%\]
\label{t.1}
\eeq
\beq
%\[
\Pi (x) = \sum_{n=0}^{15} s_n(x) \Pi (s,n,t) + \sum_{n=0}^{15}
w_n (x) \Pi (w,n,t).
%\]
\label{t.2}
\eeq
The truncated fields in (\ref{t.1}) and (\ref{t.2}) vanish smoothly at
$x=0$ and $x=20$, corresponding to the edge of the support of the left
and right-most basis function.  While more complicated boundary
condition could be used, this choice is the most straightforward to
implement, and corresponds to the projection of the wavelet basis on a
finite dimensional subspace.
    
The truncated Hamiltonian is constructed by inserting these truncated
fields in the free field Hamiltonian.  The resulting Hamiltonian is
quadratic in these fields, and has the form
\[
H =
\sum_{mn} a_{ssmn}(\lambda)  \Phi (s,m,0) \Phi (s,n,0) +
\sum_{mn} b_{ssmn}(\lambda)  \Pi (s,m,0) \Pi (s,n,0) +
\]
\[
\sum_{mn} c_{ssmn}(\lambda)  \Phi (s,m,0) \Pi (s,n,0)+
\sum_{mn} d_{ssmn}(\lambda)  \Pi (s,m,0) \Phi (s,n,0)+
\]
\[
\sum_{mn} a_{wwmn}(\lambda)  \Phi (w,m,0) \Phi (w,n,0) +
\sum_{mn} b_{wwmn}(\lambda)  \Pi (w,m,0) \Pi (w,n,0)+
\]
\[
\sum_{mn} c_{wwmn}(\lambda)  \Phi (w,m,0) \Pi (w,n,0)+
\sum_{mn} d_{wwmn}(\lambda)  \Pi (w,m,0) \Phi (w,n,0)+
\]
\[
\sum_{mn} a_{wsmn}(\lambda)  \Phi (w,m,0) \Phi (s,n,0) +
\sum_{mn} b_{wsmn}(\lambda)  \Pi (w,m,0) \Pi (s,n,0)+
\]
\[
\sum_{mn} c_{wsmn}(\lambda)  \Phi (w,m,0) \Pi (s,n,0)+
\sum_{mn} d_{wsmn}(\lambda)  \Pi (w,m,0) \Phi (s,n,0)+
\]
\[
\sum_{mn} a_{swmn}(\lambda)  \Phi (s,m,0) \Phi (w,n,0) +
\sum_{mn} b_{swmn}(\lambda)  \Pi (s,m,0) \Pi (w,n,0)+
\]
\beq
\sum_{mn} c_{swmn}(\lambda)  \Phi (s,m,0) \Pi (w,n,0)+
\sum_{mn} d_{swmn}(\lambda)  \Pi (s,m,0) \Phi (w,n,0) .
\label{t.3}
\eeq

The initial condition $(\lambda=0)$ corresponds to the original truncated
Hamiltonian, including all of the wavelet-scale coupling terms:
%\beq
\[
a_{ssmn}(0)= {1 \over 2}(\mu^2 \delta_{mn} + D_{ssmn})
\qquad
b_{ssmn}(0)= {1 \over 2} \delta_{mn}
\qquad
c_{ssmn}(0)=0
\qquad
d_{ssmn}(0)=0
\]
%\label{t.4}
%\eeq
%\beq
\[
a_{wwmn}(0)= {1 \over 2} (\mu^2 \delta_{mn} + D_{wwmn})
\qquad
b_{wwmn}(0)= {1 \over 2} \delta_{mn}
\qquad
c_{wwmn}(0)=0
\qquad
d_{wwmn}(0)=0
\]
%\label{t.5}
%\eeq
%\beq
\[
a_{wsmn}(0)= D_{wsmn}
\qquad
b_{wsmn}(0)= 0
\qquad
c_{wsmn}(0)=0
\qquad
d_{wsmn}(0)=0
\]
%\label{t.6}
%\eeq
%\beq
\[
a_{swmn}(0)= D_{swmn}
\qquad
b_{swmn}(0)= 0
\qquad
c_{swmn}(0)=0
\qquad
d_{swmn}(0)=0 .
\]
%\eeq

To test the flow equation method the mass $\mu$ was set to $1$.  The
mass sets the energy scale for the parameter $\lambda$.  An attempt to
solve the flow equation by perturbation theory did not to converge.
Convergence was achieved by solving the flow equation using the Euler
method, which uses the differential equation to step to successive
values of $\lambda$.  The step size was determined by examining the
size and number of matrix elements, to ensure that the errors remain
small.  A step size of .001 was used in our calculations.  While the
efficiency could be improved with a higher-order solution method, in
this test the Euler method was sufficient to see the that flow
equation drives the coupling term to zero.

To illustrate the evolution of the coefficients in the expansion
(\ref{t.3}) we plot the Hilbert-Schmidt norms of the non-zero
coefficients
%\beq
\[
\sqrt{\sum_{ij} a^*_{xyij}a_{xyji}} 
\qquad
\sqrt{\sum_{ij} b^*_{xyij}b_{xyji}}
\qquad
\sqrt{\sum_{ij} c^*_{xyij}c_{xyji}}
\qquad
\sqrt{\sum_{ij} d^*_{xyij}d_{xyji}} 
\]
%\eeq
as functions of $\lambda$.  

There are four types of operators, $\Phi (s,n,0),\Pi (s,n,0),\Phi
(w,m,0)$ and $\Pi (w,m,0)$ leading to 16 types of quadratic
expressions.  Figures 1-8 show the Hilbert-Schmidt norms of the
coefficients of each of the non-zero quadratic expressions as a
function of the flow parameter.

The norms of coefficients involving all scaling or all wavelet fields
evolve to non-zero values, while the norms of the coupling matrices
all evolve to zero.  The plots show that initially the size of the
coupling terms fall off very fast, but the rate of decrease slows
significantly as $\lambda$ gets larger.  For $\lambda=20$ the Hilbert
Schmidt norms of the coupling coefficients are reduced by about two
orders of magnitude from their original values.

\begin{figure}
\begin{minipage}[t]{.47\linewidth}
\centering
\includegraphics[scale=.7]{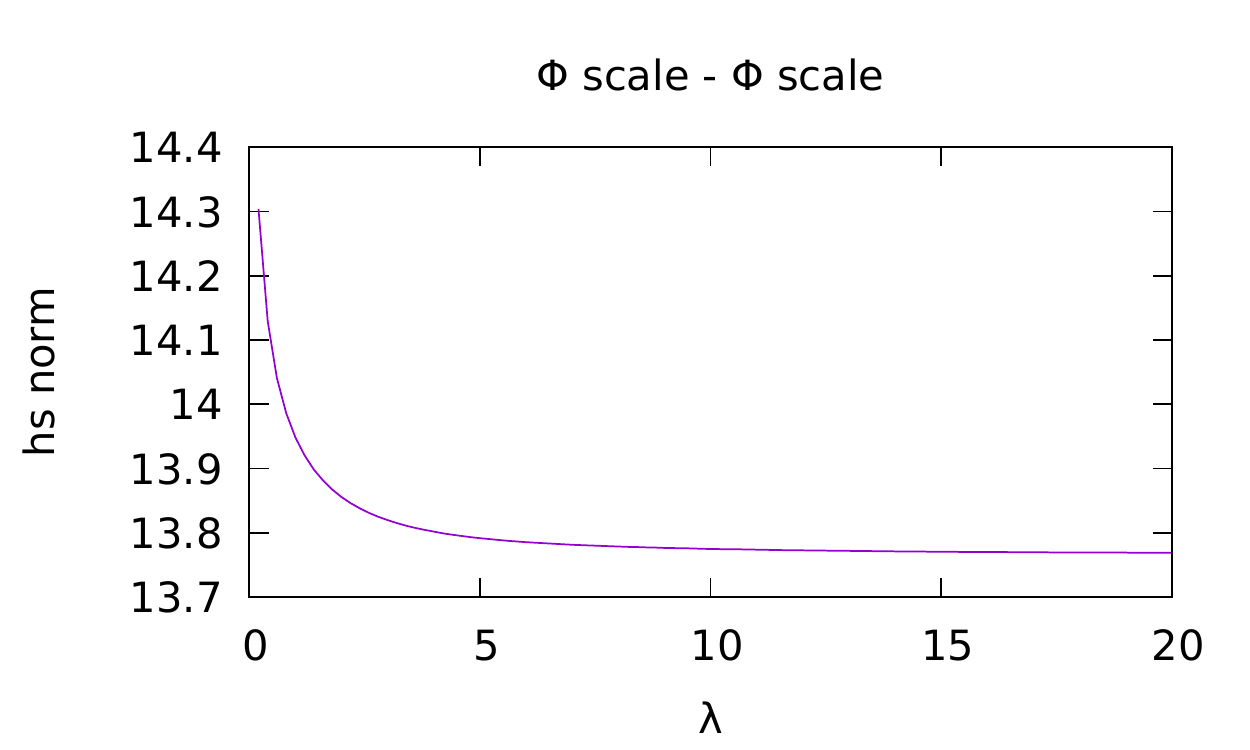}
\caption{Hilbert-Schmidt norm: $\Phi$ scale  - $\Phi$  scale}
\label{fig:1}
\end{minipage}
\begin{minipage}[t]{.47\linewidth}
\centering
\includegraphics[scale=.7]{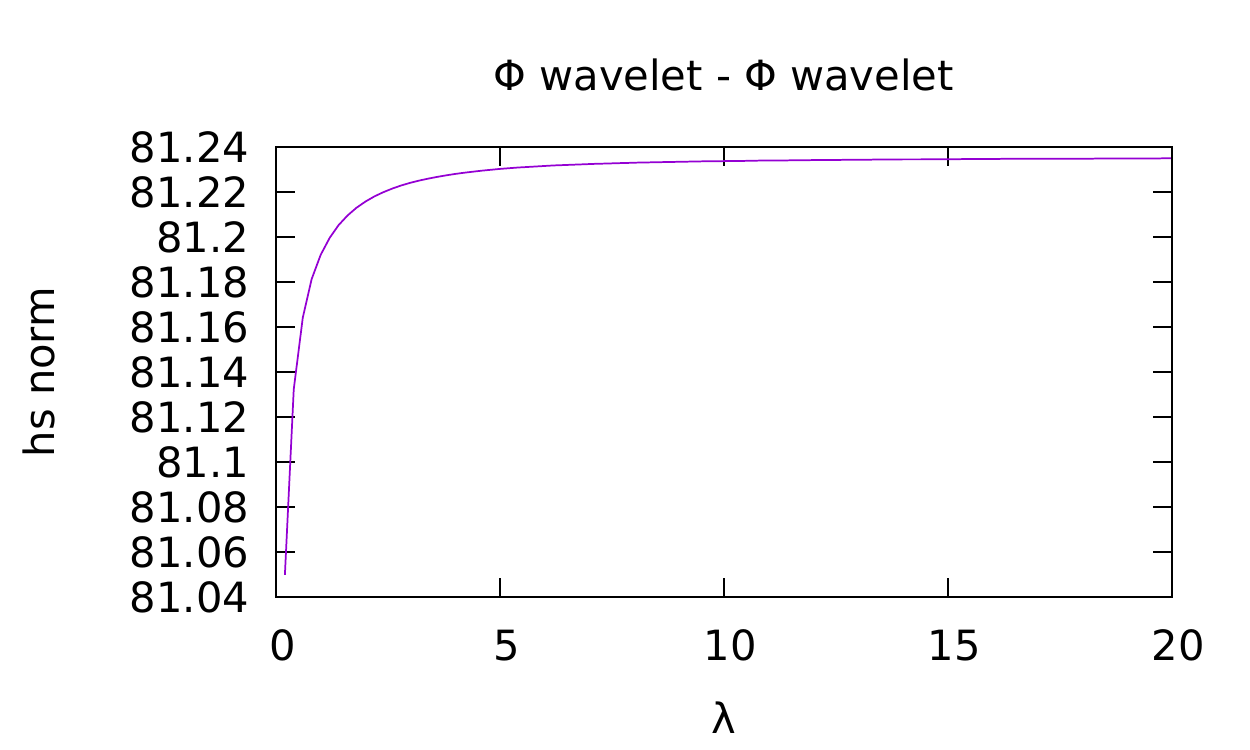}
\caption{Hilbert-Schmidt norm: $\Phi$ wavelet  - $\Phi$  wavelet}
\label{fig:2}
\end{minipage}
\end{figure}

\begin{figure}
\begin{minipage}[t]{.47\linewidth}
\centering
\includegraphics[scale=.7]{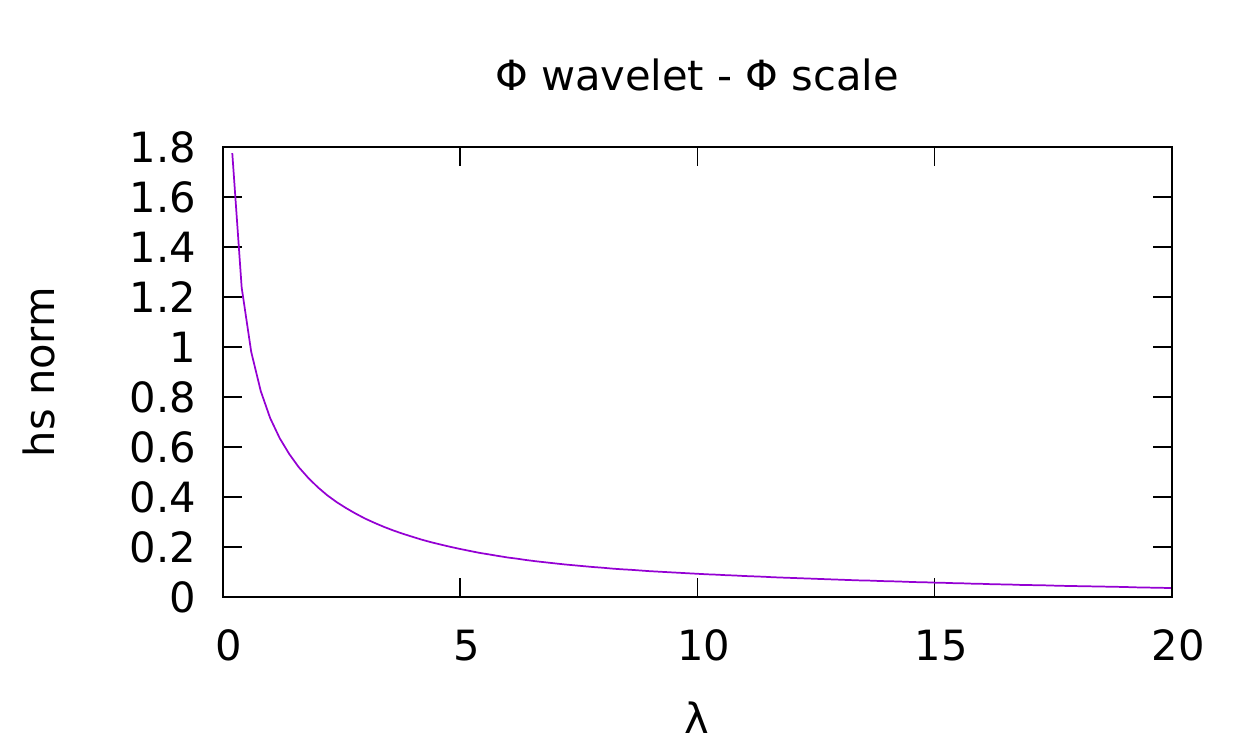}
\caption{Hilbert-Schmidt norm: $\Phi$ scale  - $\Phi$  scale}
\label{fig:3}
\end{minipage}
\begin{minipage}[t]{.47\linewidth}
\centering
\includegraphics[scale=.7]{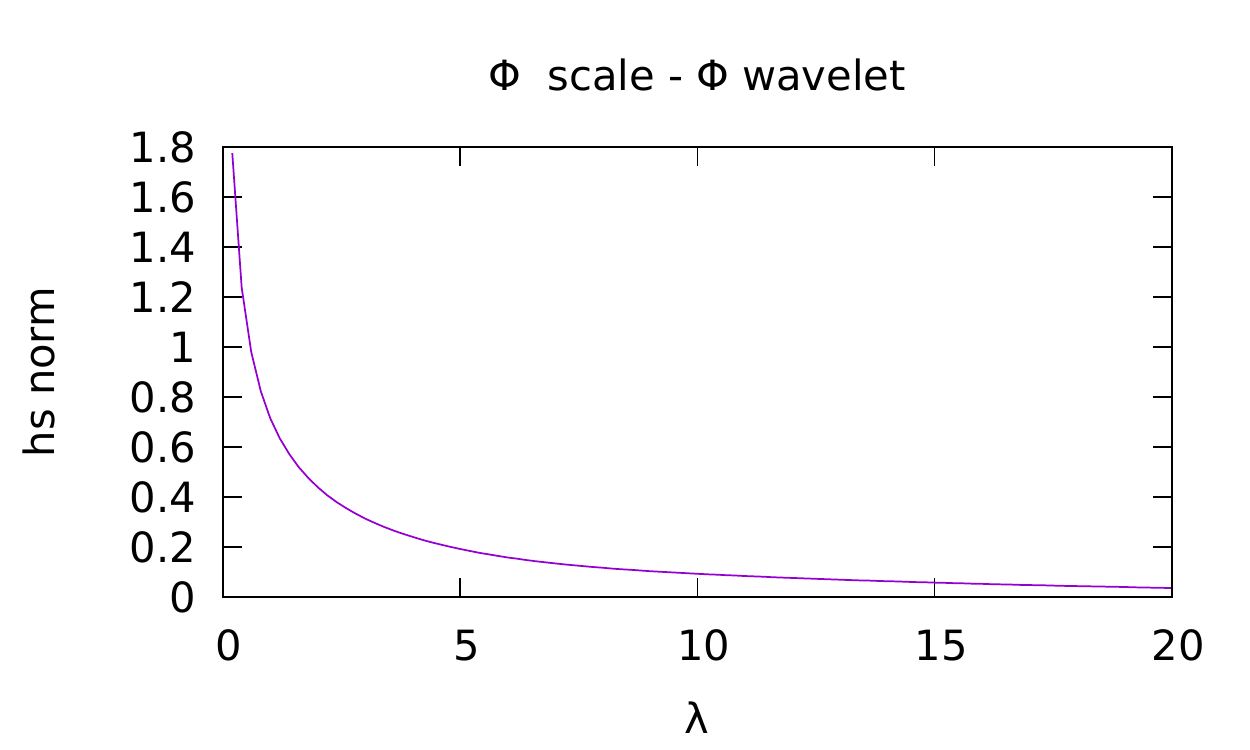}
\caption{Hilbert-Schmidt norm: $\Phi$ scale  - $\Phi$  wavelet}
\label{fig:4}
\end{minipage}
\end{figure}

\begin{figure}
\begin{minipage}[t]{.47\linewidth}
\centering
\includegraphics[scale=.7]{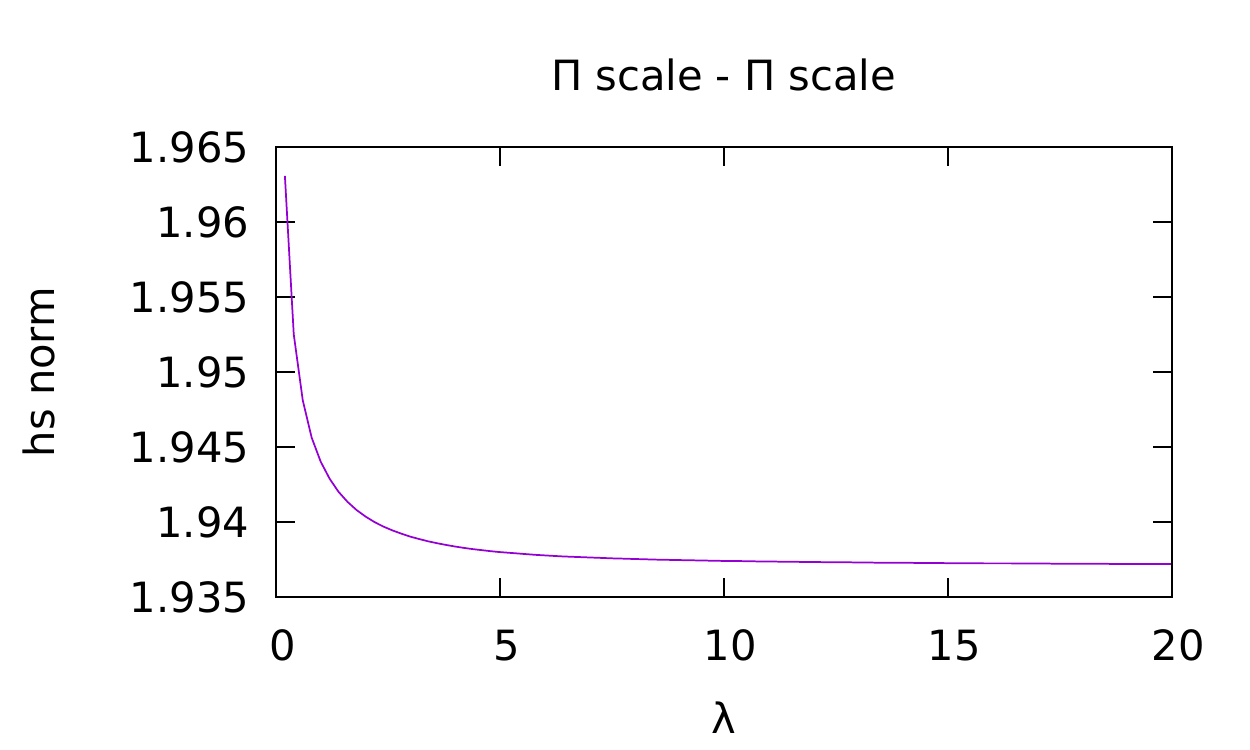}
\caption{Hilbert-Schmidt norm: $\Pi$ scale  - $\Pi$  scale}
\label{fig:5}
\end{minipage}
\begin{minipage}[t]{.47\linewidth}
\centering
\includegraphics[scale=.7]{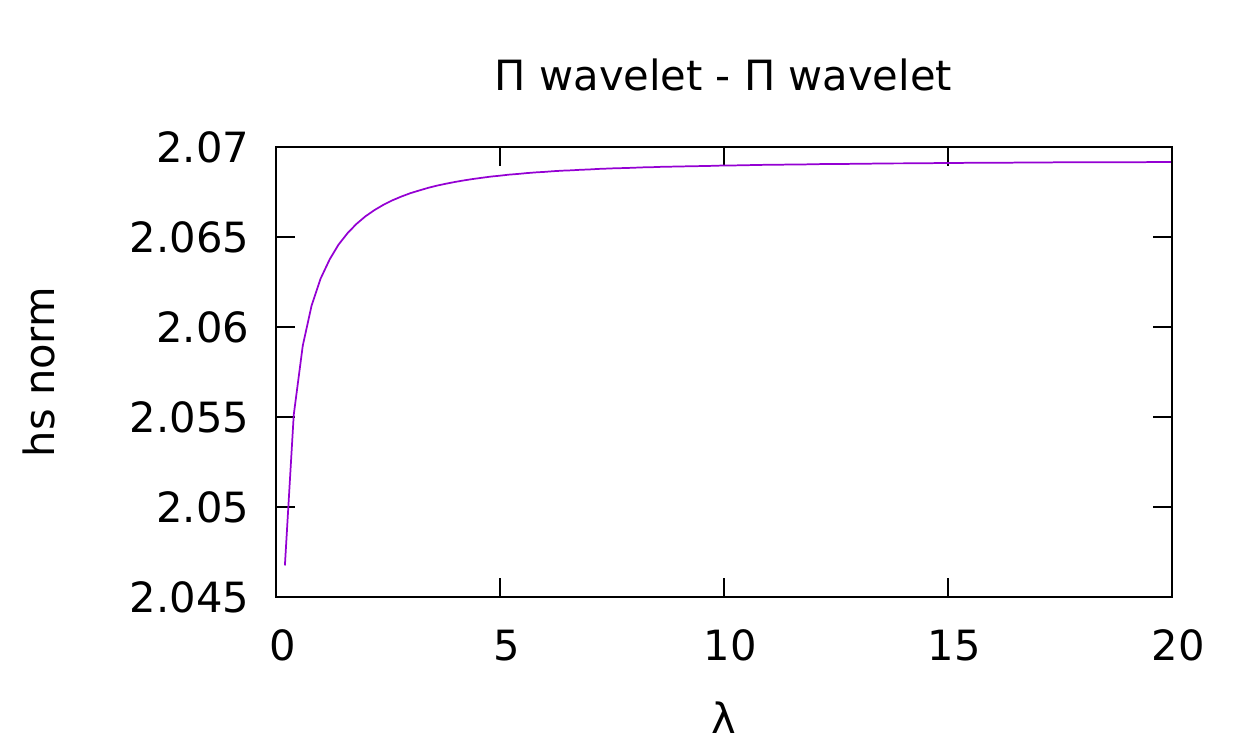}
\caption{Hilbert-Schmidt norm: $\Pi$ wavelet  - $\Pi$  wavelet}
\label{fig:6}
\end{minipage}
\end{figure}

\begin{figure}
\begin{minipage}[t]{.47\linewidth}
\centering
\includegraphics[scale=.7]{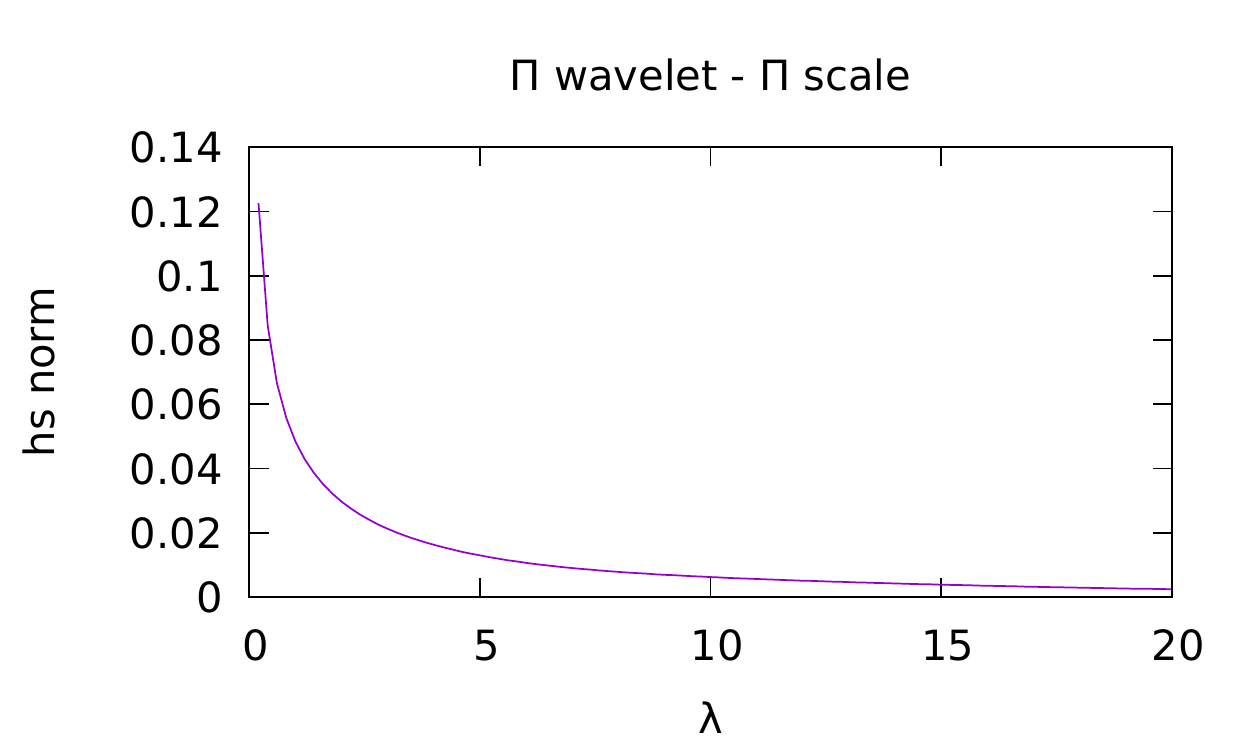}
\caption{Hilbert-Schmidt norm: $\Pi$ wavelet  - $\Pi$  scale}
\label{fig:7}
\end{minipage}
\begin{minipage}[t]{.47\linewidth}
\centering
\includegraphics[scale=.7]{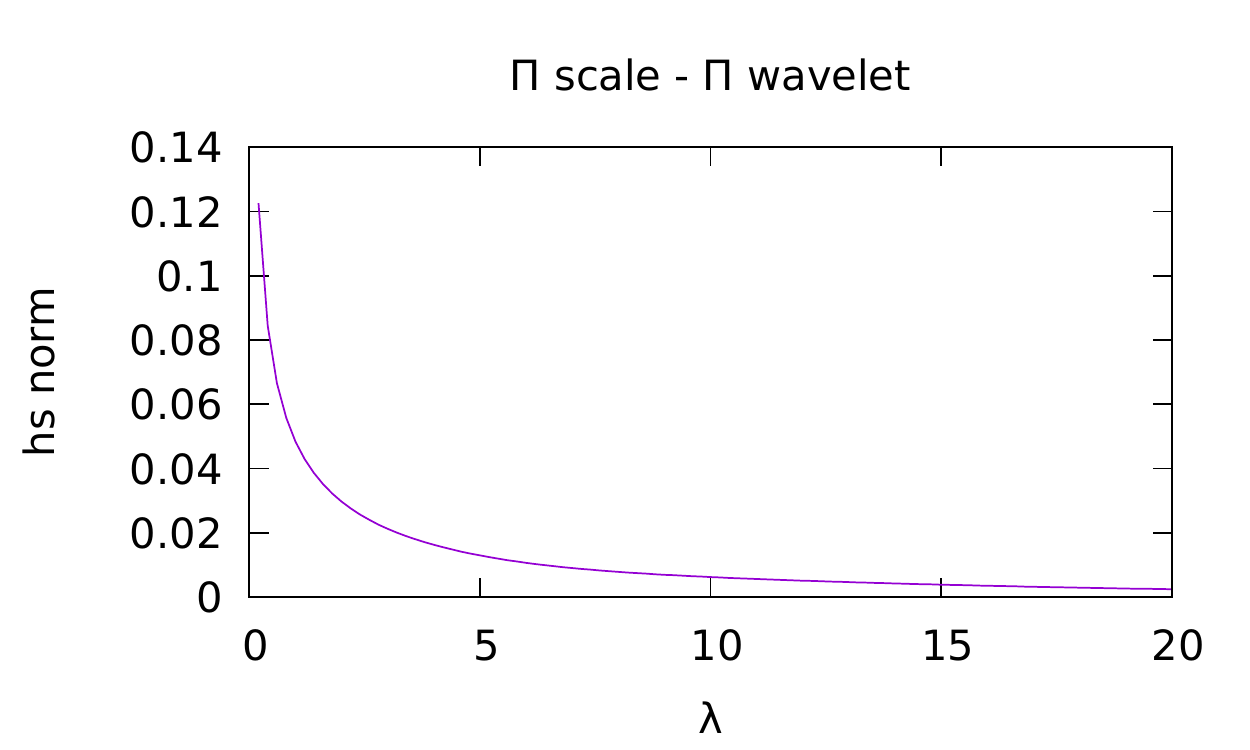}
\caption{Hilbert-Schmidt norm: $\Pi$ scale  - $\Pi$  wavelet}
\label{fig:8}
\end{minipage}
\end{figure}

\begin{figure}
\begin{minipage}[t]{.47\linewidth}
\centering
\includegraphics[angle=000,scale=.45]{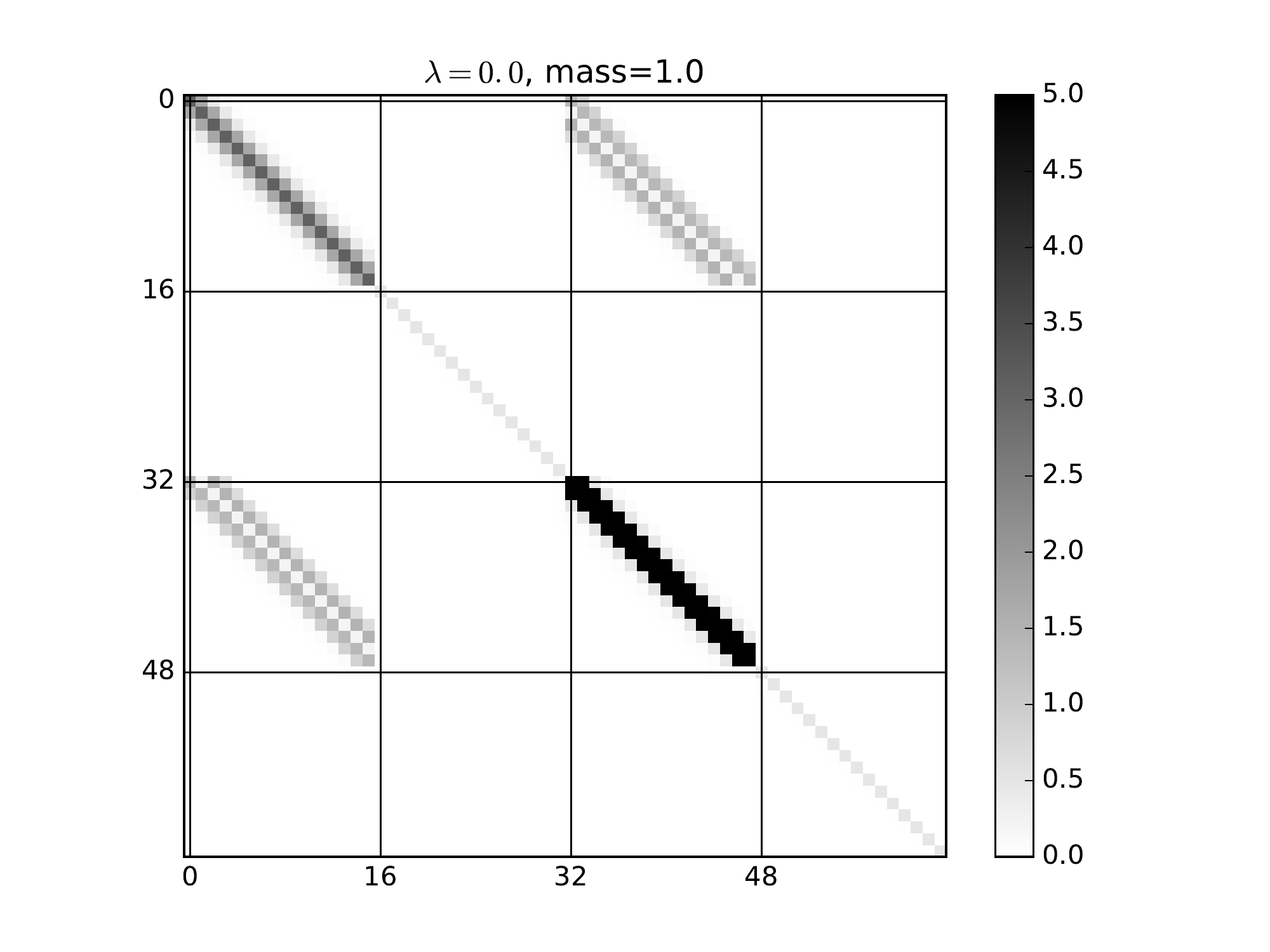}
\caption{Full matrix, $\lambda$=0, mass=1}
\label{fig:8}
\end{minipage}
\begin{minipage}[t]{.5\linewidth}
\centering
\includegraphics[angle=000,scale=.45]{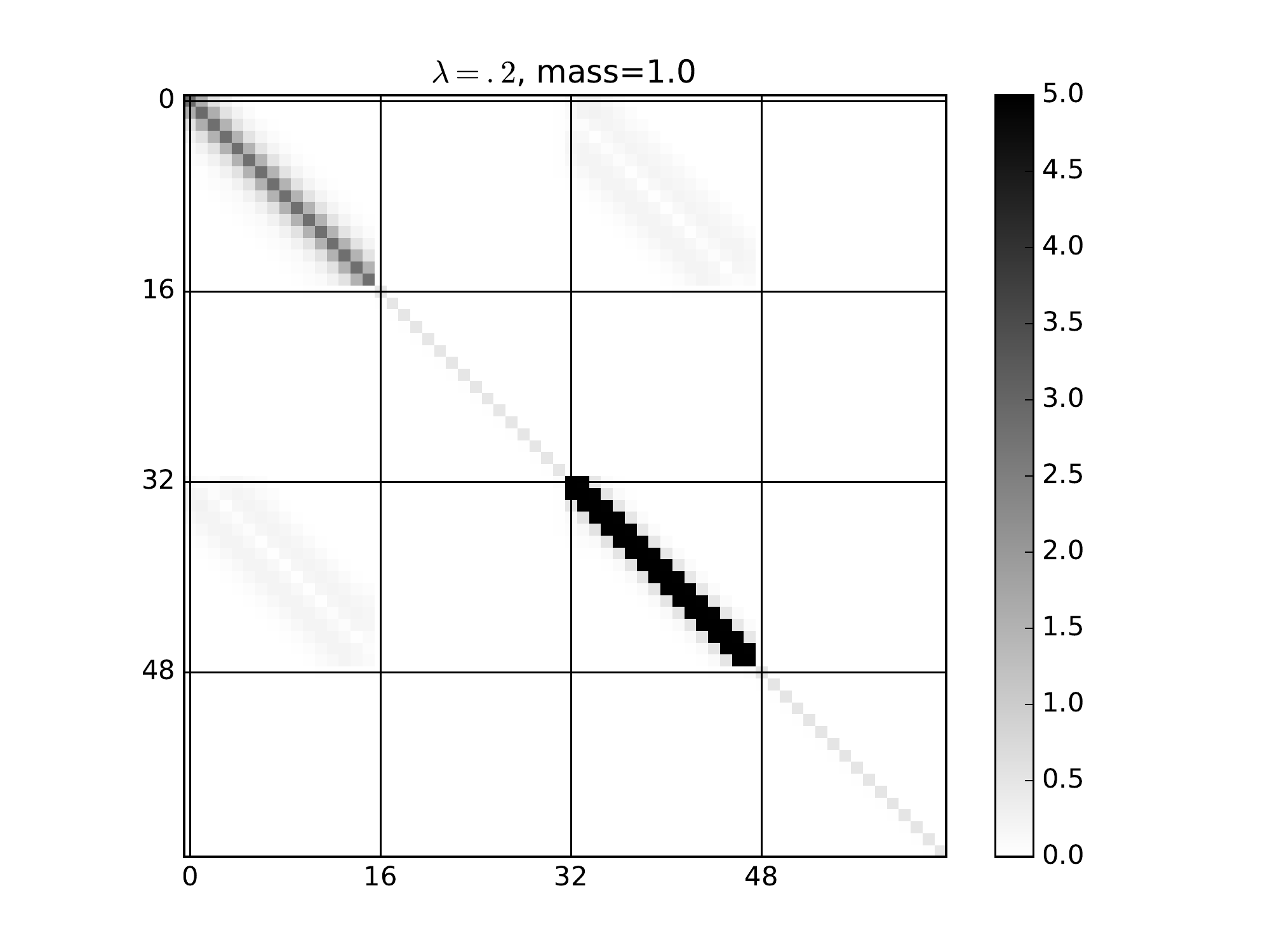}
\caption{Full matrix, $\lambda$=0.2, mass=1}
\label{fig:9}
\end{minipage}
\end{figure}

\begin{figure}
\begin{minipage}[t]{.47\linewidth}
\centering
\includegraphics[angle=000,scale=.45]{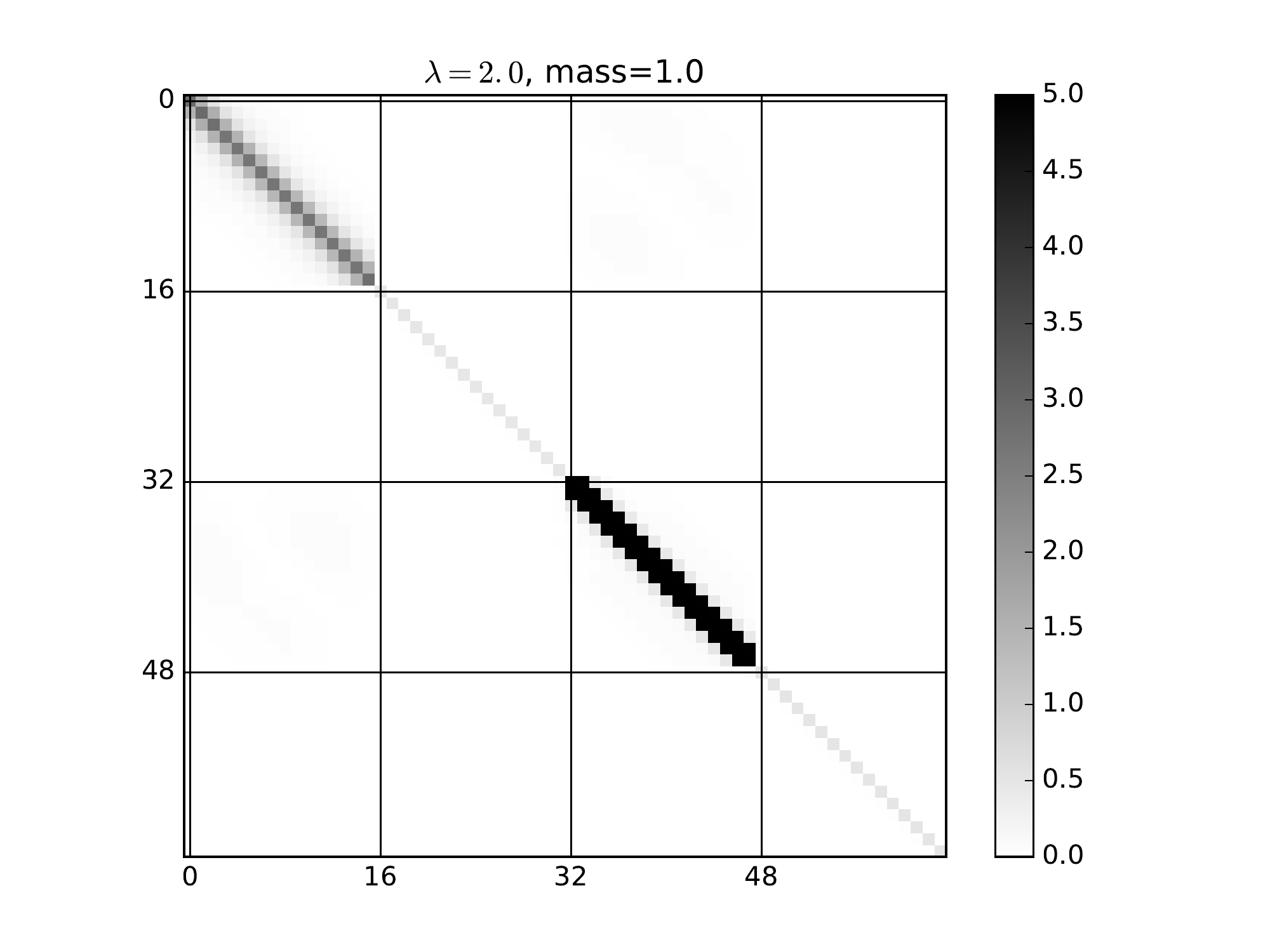}
\caption{Full matrix, $\lambda$=2.0, mass=1}
\label{fig:10}
\end{minipage}
\begin{minipage}[t]{.5\linewidth}
\centering
\includegraphics[angle=000,scale=.45]{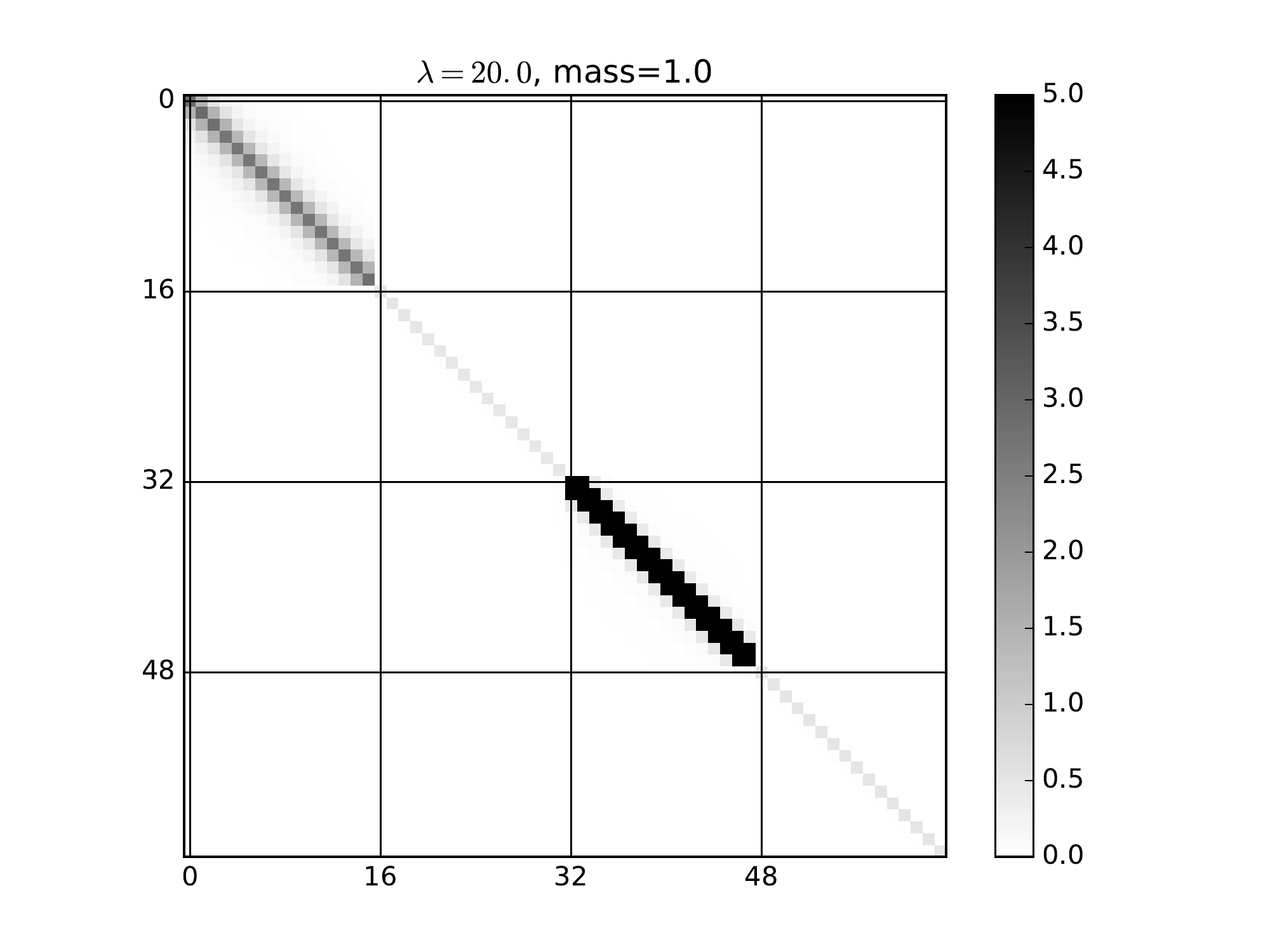}
\caption{Full matrix, $\lambda$=20.0, mass=1}
\label{fig:11}
\end{minipage}
\end{figure}

The Hilbert-Schmidt norms of the coefficients are dominated by the
largest matrix elements.  It is also useful to understand how the 
individual matrix coefficients converge.

Figures 9-12 provide a graphical representation of
the coefficient matrices for different values of $\lambda$.  
The figures should be viewed as a
montage of sixteen 16$\times$16 matrices.  Indices 0-15 correspond
to $\Phi(s)$, 16-31 correspond to $\Pi(s)$, 32-47 correspond to $\Phi(w)$ and
48-63 correspond to $\Pi(w)$.  The gray scale shows the size of the 
coefficients of the quadratic expressions  
in the Hamiltonian as a function of $\lambda$.

The four figures correspond to different values of the flow parameter
$\lambda=0$, $\lambda=.2$, $\lambda=2$ and $\lambda =20$.

Figure 9 represents the initial values.  The two narrow diagonal bands
in the 16-31 and 48-63 blocks represent the coefficients $b_{ssmn}(0)$
 and $b_{wwmn}(0)$ respectively.  The fatter diagonal bands in the
upper left hand part of this figure are associated with the
scale-scale and wavelet-wavelet derivative terms.  They are almost
diagonal because the matrices $D_{x;mn}$ only couple neighboring
degrees of freedom.

The terms above and below the diagonal are the coefficients of the
scale-wavelet and wavelet-scale derivative terms, $a_{swmn}(0)$ and
$a_{wsmn}(0)$.  These are responsible for the coupling of the two
scales and are the terms that the flow equation is designed to
suppress.

Figure 10 shows the value of these coefficients for $\lambda=.2$.
For this value of $\lambda$ the coupling terms have become 
smaller and more non-local.  This is because repeated applications of 
the derivative matrix widens the support of the degrees of freedom that 
are coupled together. 

Figure 11 shows that by $\lambda=2$ the scale coupling terms have essentially
disappeared.

Figure 12 show that integrating the flow equation out to $\lambda =20$
does not lead to any big changes.  This is consistent with the
behavior shown in figures 1-8, that the exponential suppression slows
as $\lambda$ is increased.  It is worth noting that the width of the
diagonal band in the uncoupled Hamiltonian at $\lambda =20$ is about
the same size as the width of the corresponding band in the original
Hamiltonian.  This shows that at least for this example the flow
equation approximately preserves the local nature of the truncated theory.

The figures support the contention that flow equation methods
can be successfully applied to separate scales in
wavelet discretized field theory.
  
%An interacting theory will have a Hamiltonian with a similar band 
%structure.  In addition to bands associated with the derivative terms,
%there will be new tensor structures associated with field 
%couplings.

For any value of the flow parameter the flow equation generates a new
equivalent Hamiltonian.  Ignoring the small terms that couple the
wavelet to the scaling function fields, the Hamiltonian becomes a sum
of two commuting operators that have degrees of freedom associated
with different scale degrees of freedom, but each operator includes
the effects of the eliminated degrees that appear in the other
operator.

In the free field case the Hamiltonians are quadratic 
functions of the canonical fields for any value of the
flow parameter.  For sufficiently large values of the
flow parameter the Hamiltonian becomes sums of two Hamiltonians involving
different scale degrees of freedom.

It is still necessary to solve for the vacuum  and solve the field
equations of the truncated theory.  In this
case there are two independent systems of field equations using the
decoupled Hamiltonians:
%\beq
\[
\dot{\Phi}(s,n,t) = i [H_s(\lambda), {\Phi}(s,n,t)]  
\qquad
\dot{\Pi}(s,n,t) = i [H_s(\lambda), {\Pi}(s,n,t)]  
\]
%\eeq
and
%\beq
\[
\dot{\Phi}(w,n,t) = i [H_w(\lambda), {\Phi}(w,n,t)]  
\qquad
\dot{\Pi}(w,n,t) = i [H_w(\lambda), {\Pi}(w,n,t)]  
\]
%\eeq
The effective Hamiltonian $H_s(\lambda)$ is the effective
Hamiltonian with the relevant (scaling-function) degrees of freedom.
For the free field, the dynamics corresponds to a set of
coupled Harmonic oscillators.  This can be used to solve the 
Heisenberg equations, compute the vacuum, and compute
approximate correlation functions for the coarse scale block
Hamiltonian.  This is discussed in more detail in the next section.

%give the time dependence of the discrete
%field operators and can be used with the expansions (\ref{f.11}-\ref{f.12}) 
%to compute approximate spacetime correlation functions.

%While this problem is solvable for the
%free field case, the same approach can be applied to the interacting
%case.  The only difference is that $H_s(\lambda)$ will be a more
%complicated operator.

\section{analysis}

The results of the previous section show that in the free field case
that flow equation methods can be used to approximately block
diagonalize the truncated Hamiltonian by scale.  To better understand
properties of the solution is it helpful to first consider properties
of the exact solution of the truncated equations and how they respond
to changes in volume and resolution truncations.  Volume truncations
change the number of canonical pairs of fields while resolution
truncations only change the overlap matrices of the spatial
derivatives (\ref{f.18})-(\ref{f.20}).  In general the initial
truncated Hamiltonian can be expressed in matrix form as
\[
H = {1 \over 2} 
[
(\Pi^s ,\Pi^w) 
\left (
\begin{array}{cc}
I_s & 0\\
0 & I_w 
\end{array}
\right ) 
\left (
\begin{array}{c}
\Pi^s \\
\Pi^w  
\end{array}
\right ) +
(\Phi^s ,\Phi^w) 
\left (
\begin{array}{cc}
\mu^2I+D_{s} & D_{sw}\\
D_{ws}  & \mu^2I + D_{w} 
\end{array}
\right ) 
\left (
\begin{array}{c}
\Phi^s \\
\Phi^w  
\end{array}
\right ) 
]
\]
where the upper components represent the scaling function fields and the
lower components represent the wavelet fields.  Because the matrix
\[
M:= 
\left (
\begin{array}{cc}
\mu^2I+D_{s} & D_{sw}\\
D_{ws}  & \mu^2I + D_{w} 
\end{array}
\right ) 
\]
is a real symmetric matrix it can be diagonalized by a real orthogonal 
matrix $O$:
\beq
O^t M O 
%\left (
%\begin{array}{cc}
%Im^2 + D^{ss} & D^{sw}0\\
%D^{ws} & Im^2 + D^{ww}  
%\end{array}
%\right ) 
%O
=
\left (
\begin{array}{cc}
m^s & 0\\
0 & m^w  
\end{array}
\right )
\label{an.1}
\eeq
where $m^s$ and $m^w$ are diagonal matrices consisting 
of eigenvalues of the
matrix $M$.

Transformed discrete fields are defined by
\[
\left (
\begin{array}{c}
\Phi^{\prime s} \\
\Phi^{\prime w}  
\end{array}
\right )  := 
O
\left (
\begin{array}{c}
\Phi^s \\
\Phi^w  
\end{array}
\right ) 
\qquad \mbox{and} \qquad
\left (
\begin{array}{c}
\Pi^{\prime s} \\
\Pi^{\prime w}  
\end{array}
\right )  := 
O \left (
\begin{array}{c}
\Pi^s \\
\Pi^w  
\end{array}
\right ) .
\]
The orthogonality of $O$ implies that the transformed fields satisfy
canonical commutation relations.  It follows from the Stone-von
Neumann uniqueness theorem that this transformation of the field
operators can be implemented by a unitary transformation $W$:
\[
W 
\left (
\begin{array}{c}
\Phi^s \\
\Phi^w  
\end{array}
\right ) 
W^{\dagger} = 
\left (
\begin{array}{c}
\Phi^{\prime s} \\
\Phi^{\prime w}  
\end{array}
\right )  = 
O \left (
\begin{array}{c}
\Phi^s \\
\Phi^w  
\end{array}
\right ) 
\qquad
\mbox{and}
\qquad
W 
\left (
\begin{array}{c}
\Pi^s \\
\Pi^w  
\end{array}
\right ) 
W^{\dagger} = 
\left ( 
\begin{array}{c}
\Pi^{\prime s} \\
\Pi^{\prime w}  
\end{array}
\right )  = 
O \left (
\begin{array}{c}
\Pi^s \\
\Pi^w  
\end{array}
\right ) . 
\]
If this transformation is applied to the truncated Hamiltonian it is
transformed into the sum of uncoupled harmonic oscillator
Hamiltonians, where the squares of the oscillator frequencies are
exactly the eigenvalues of the matrix $M$:
\[
H'= UHU^{\dagger} = {1 \over 2} 
[
(\Pi^{ s} ,\Pi^{ w})
\left (
\begin{array}{cc}
I & 0\\
0 & I 
\end{array}
\right ) 
\left (
\begin{array}{c}
\Pi^{ s} \\
\Pi^{ w}  
\end{array}
\right ) +
(\Phi^{ s} ,\Phi^{w}) 
\left (
\begin{array}{cc}
m^{s} & 0\\
0  & m^{w} 
\end{array}
\right ) 
\left (
\begin{array}{c}
\Phi^{ s} \\
\Phi^{ w}  
\end{array}
\right ) 
]
\]
or 
\[
H = {1\over 2} \sum_n \left (\Pi (s,n,0)\Pi (s,n,0) +
m^s_{n} \Phi (s,n,0) \Phi (s,n,0) \right ) +
{1\over 2} \sum_n \left (  \Pi (w,n,0) \Pi (w,n,0) + 
m^w_{n} \Phi (w,n,0) \Phi (w,n,0)\right ). 
\]
The ground state of the truncated Hamiltonian is the state annihilated
by the annihilation operators
\beq
a^s_{n} :=  {1 \over \sqrt{2}m_{n}^{s 1/4}}\sum_j
(\sqrt{m^s_{n}}\Phi (s,n,0) + i\Pi (s,n,0)) 
\eeq
\beq
a^w_{n} :=  {1 \over \sqrt{2}m_{n}^{w 1/4}}\sum_j
(\sqrt{m^w_{n}}\Phi (w,n,0) + i\Pi (w,n,0)) 
\eeq
where $m^s_n$ and $m^w_n$ are the eigenvalues of the diagonal matrices
$m^s$ and $m^w$.  It follows that the unitary operator $U$ does a
complete diagonalization that separates different scale degrees of
freedom.  The transformation $O$ is not unique, since permutations of
the columns of $O$ permute the eigenvalues. This means that the
identification of a given oscillator frequency with a wavelet or
scaling function degree of freedom depends on the choice of $O$.

The advantage of this representation is that it can be used to
understand the behavior of the truncated Hamiltonian with respect to
changes in volume and resolution and well as the role of the mass
parameter.  The key observation is that the truncated system is
equivalent to a set of uncoupled harmonic oscillators where (1) the
number of oscillators is proportional to the cutoff volume and the
oscillator frequencies are square roots of the eigenvalues of the
matrix $M$.  The matrix $M$ is $\mu^2 I + D$, where $D$ is a positive
symmetric matrix with eigenvalues $d_i$.  The matrix $D$ is the only
quantity in the truncated Hamiltonian that changes under scale
changes.  If we double the resolution, $D \to 4 D$.  This means that
the eigenvalues of $M$ have the form $m_i= \mu^2 + d_i$ and under a
change of resolution by a factor of 2 they become $m_i \to \mu^2 + 4
d_i$.  This means that doubling the resolution increases the
separation between the squares of the oscillator frequencies by a
factor of 4.  The other property of $D$ is that, up to boundary terms,
it is translationally invariant.  If we think of it as representing
the kinetic energy of particles in a box, we expect that doubling the
box size will introduce new modes with half of the frequency.  Thus,
while increasing the resolution increases the separation between
oscillator frequencies, if the volume is increased, new lower
frequency modes are added that fill in the gaps generated by
increasing the resolution.  The mass $\mu$ provides a lower bound for
the oscillator frequencies.  Since spectrum of the exact free field
Hamiltonian is continuous, in order to get a continuum limit, it is
necessary to simultaneously increase both the volume and resolution in
a manner that the separation between adjacent normal mode frequencies
vanishes.

The resolution for a fixed scale can also be improved by increasing
the order $K$ of the scaling function-wavelet basis.  For a
fixed scale, basis functions with higher values of $K$ can locally
pointwise represent higher degree polynomials than the $K=3$ basis
functions \cite{beylkin4}.  The cost is that the basis functions have
larger support and are thus less local for a given level of
truncation.  The improvement in efficiency by increasing $K$ for a
fixed scale wavelet-truncated free field theory was demonstrated in
\cite{singh}.  

It is also interesting to note that the scaling properties of
Hamiltonian can be misleading.  Specifically, for the free field
Hamiltonian, in the infinite resolution limit it looks like the matrix
$D \to 4^kD$ which should eventually dominate the fixed mass for large
$k$.  However it is clear that the mass cannot be ignored, since it
fixes the minimum value of the energies for any scale.  This suggest
that it might be more useful to consider properties correlation
functions under change of volume and scale, since these quantities
also depend on the vacuum.  These issues will clearly become more
complex for interacting theories.  The correlation function (Wightman
function) for the discrete field is
\[
\langle 0 \vert \Phi (x,t_x) \Phi (y,t_y) \vert 0 \rangle = 
\langle 0' \vert \Phi' (x) \Phi' (y) \vert 0' \rangle = 
\sum_{mnk} s_n(x) s_m (y) O_{nk} O_{mk}  
{1 \over 2 \sqrt{m_k}}e^{i m_k (t_y-t_x)} .
\]
These are discussed in \cite{singh}. 

The discussion above does not directly apply to the block diagonal
Hamiltonian constructed by the flow equation, however the transformation
$U(\lambda)$ generated by the flow cannot change the spectrum of 
the Hamiltonian.  It can only determine which of the exact
eigenvalues get assigned to each of the two blocks.  The scaling
properties of the eigenvalues are the same as in the exact case.

We can understand what happens in the flow equation case.  For mass
$\mu=1$ and $\lambda =20$ the coefficients $b_{ssmn}$ and $b_{wwmn}$
are within about 2\% of their initial values.  This means that the
squares of the normal mode frequencies of the scaling block diagonal
Hamiltonian are approximately eigenvalues of the matrix
$2 a_{ssmn}(\lambda=20)$.  These can be compared to the corresponding
normal model frequencies of the full truncated Hamiltonian.

In table 2 we display the squares of the $\lambda=20$ normal mode
frequencies in column 1, the squares of the normal mode frequencies of
the Hamiltonian obtained by simply throwing away the wavelet degrees
of freedom without eliminating them in column 2, and the squares of
the normal mode frequencies of the full truncated Hamiltonian in
columns 3 and 4.  Inspection of this table shows that the frequencies
in the first column are approximately equal to the frequencies in the
third column.  This shows that the flow equation block diagonailzes
the Hamiltonian into a block with the 16 lowest frequency modes (the
coarse scale block) and another block with the 16 highest frequency
modes (fine scale block).  The properties of these matrices under
change of scale or resolution follow from the behavior of the normal
mode frequencies - increasing the volume adds more low frequency
modes, while increasing the resolution increases the separation
between the different normal mode frequencies.  The mass sets the
lowest normal mode frequency.

\begin{table}[t]
\caption{Normal mode frequencies}
\begin{tabular}{|l|l|l|l|}
\hline		
$\lambda=20, \mu=1$& truncated & exact 1:16 & exact 17:32 \\
\hline
1.037e+00 &  1.037e+00 &  1.041e+00 &  1.665e+01 \\   
1.145e+00 &  1.146e+00 &  1.153e+00 &  1.925e+01 \\
1.326e+00 &  1.333e+00 &  1.340e+00 &  2.208e+01 \\
1.583e+00 &  1.609e+00 &  1.604e+00 &  2.512e+01 \\
1.919e+00 &  1.995e+00 &  1.947e+00 &  2.834e+01 \\
2.341e+00 &  2.525e+00 &  2.373e+00 &  3.167e+01 \\
2.861e+00 &  3.236e+00 &  2.890e+00 &  3.507e+01 \\
3.493e+00 &  4.161e+00 &  3.508e+00 &  3.846e+01 \\
4.263e+00 &  5.317e+00 &  4.243e+00 &  4.178e+01 \\
5.201e+00 &  6.689e+00 &  5.112e+00 &  4.495e+01 \\
6.346e+00 &  8.232e+00 &  6.134e+00 &  4.789e+01 \\
7.722e+00 &  9.859e+00 &  7.332e+00 &  5.053e+01 \\
9.309e+00 &  1.145e+01 &  8.729e+00 &  5.279e+01 \\
1.102e+01 &  1.289e+01 &  1.034e+01 &  5.462e+01 \\
1.274e+01 &  1.403e+01 &  1.219e+01 &  5.597e+01 \\
1.435e+01 &  1.476e+01 &  1.429e+01 &  5.679e+01 \\
\hline
\end{tabular}
\label{omega}
\end{table}

In this work only two scales that differ by a factor of two were
considered.  Flow equation methods easily generalize to treat
truncated theories with many different scales.  While the goal of this
paper was to construct an equivalent effective theory by eliminating
fine resolution degrees of freedom that are not explicitly needed, the
calculation is equivalent to either truncating the Hamiltonian using a
fixed fine scale truncated Hamiltonian of the form (\ref{f.15}) or using the
equivalence (\ref{b.12}) of the fine scale truncated Hamiltonian to
the multiscale truncated Hamiltonian (\ref{f.14}).  This is a
canonical transformation that can be realized unitarily.  In this work
the fine and coarse scale degrees of freedom are decoupled.  In
\cite{singh} the other two equivalent representations are used.  They
interpret the relation between the fine scale scaling-function
representation and the multiscale scaling-function-wavelet
representation,  which is given by the wavelet transform, as an
explicit example of an ADS-CFT-like bulk boundary duality, where the
multiple scales in the multiscale Hamiltonian (\ref{f.14}) represent
the discretization of an extra dimension.  For the free field example,
at a given level of fine scale and volume truncation, all three
Hamiltonians represent coupled oscillators with the same normal mode
frequencies.

Each of these representations have different advantages.  The
representation discussed in this work is focused on developing a
formulation of the theory using only degrees of freedom on one coarse
scale, which would ideally be identified with a physical scale.  The
multiscale representation has the advantage that it initially leads to
a discrete representation of the exact (untruncated) theory that can
be subsequently truncated.  The direct fine scale truncations are
similar to lattice truncations.  While they are never exact, the
scaling properties of the basis functions make this representation the
most natural one to study the limit to an arbitrarily fine resolution.
This is done explicitly in \cite{singh} where the truncated
Daubechies' free-field scaling-wavelet vacuum correlation functions in
both the massive and massless case are shown to converge to the
corresponding free-field Wightman functions in \cite{singh}
 
The one thing that was not discussed is what happens to the numerical
methods when the mass is changed.  In this free field example the mass
fixes the lower limit of the spectrum and also fixes the dimensions of
the flow parameter.  One might expect that since the exponential fall
off in (\ref{feq.21}) is determined by the separation of the evolved
normal mode eigenvalues of the block diagonal part of the Hamiltonian,
that the convergence of the flow equation will not be significantly
affected by changing the mass.  To test this we solved the flow
equation for $\mu^2=0$ and $\mu^2=16$ for $\lambda =.2$ and
$\lambda=2.$.  The results are shown in figures (13) and (14).  These
figures look very similar to the matrices in figures (10) and (11) for
$\mu^2=1$.  This suggest that the flow equation has no special
difficulties in treating truncated theories even when $m=0$.

\begin{figure}
\begin{minipage}[t]{.47\linewidth}
\centering
\includegraphics[angle=000,scale=.45]{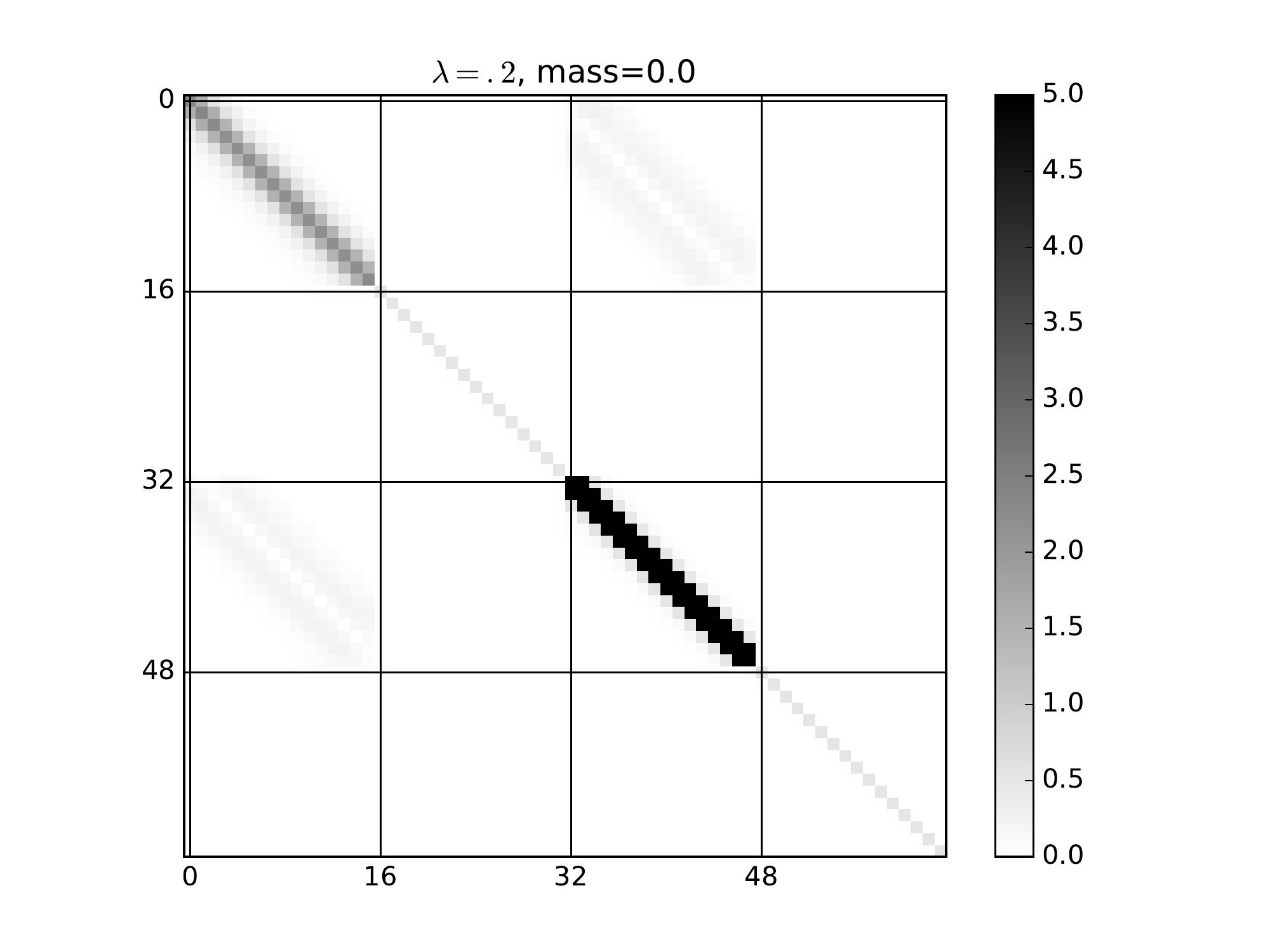}
\caption{Full matrix, $\lambda$=.2, mass=0}
\label{fig:12}
\end{minipage}
\begin{minipage}[t]{.5\linewidth}
\centering
\includegraphics[angle=000,scale=.45]{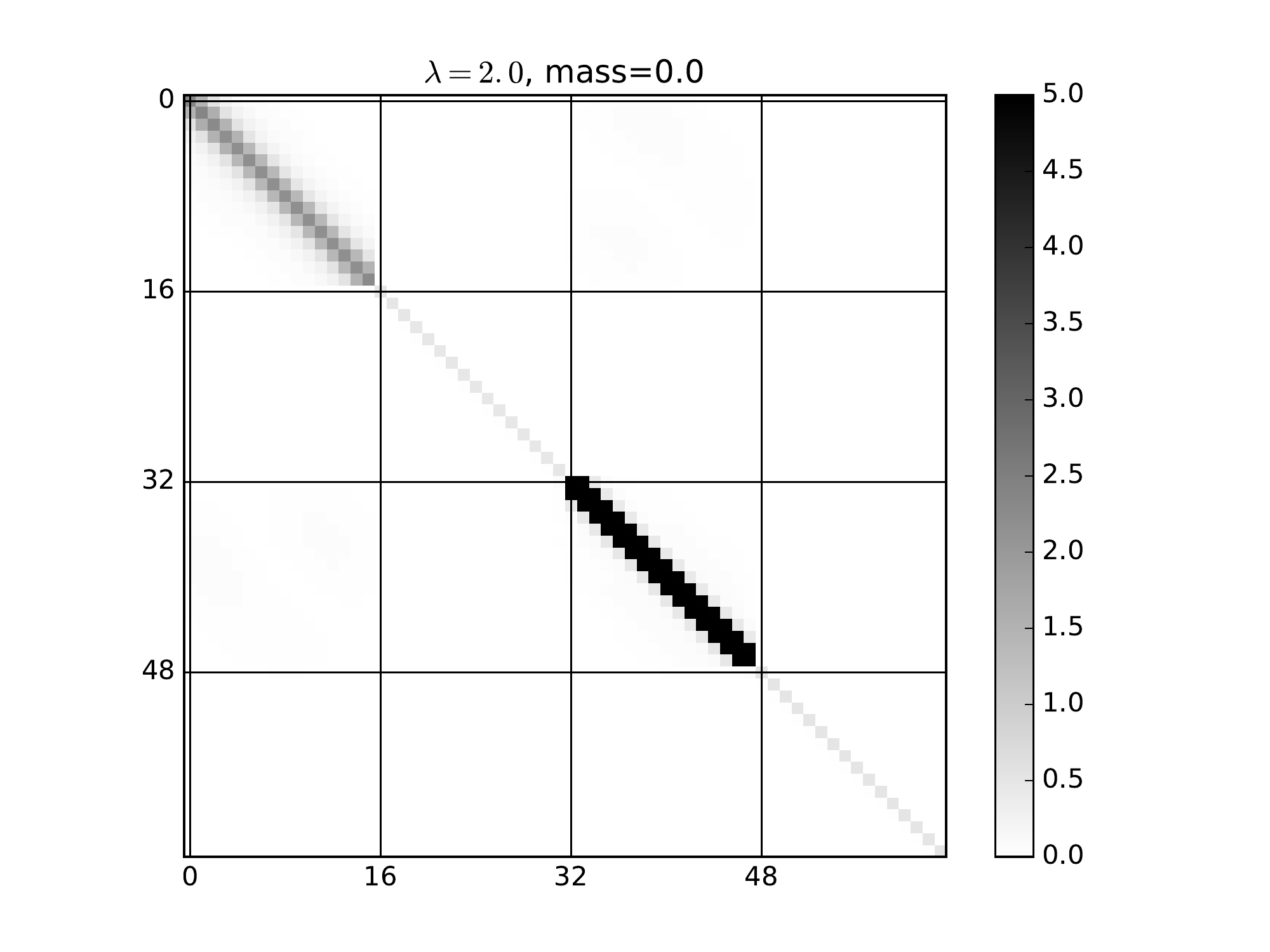}
\caption{Full matrix, $\lambda$=2.0, mass=0}
\label{fig:13}
\end{minipage}
\end{figure}

\begin{figure}
\begin{minipage}[t]{.47\linewidth}
\centering
\includegraphics[angle=000,scale=.45]{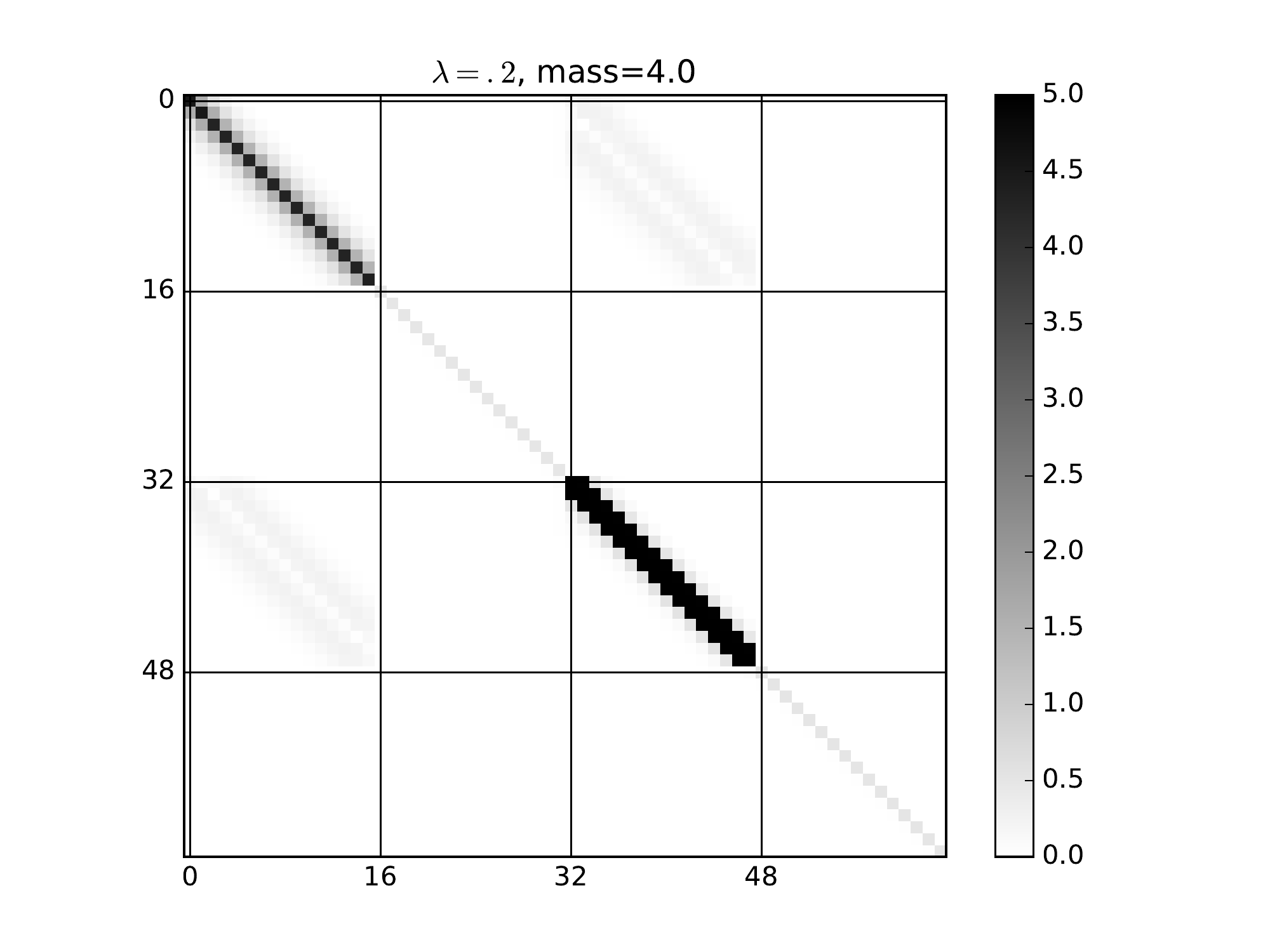}
\caption{Full matrix, $\lambda$=.2, mass=4}
\label{fig:14}
\end{minipage}
\begin{minipage}[t]{.5\linewidth}
\centering
\includegraphics[angle=000,scale=.45]{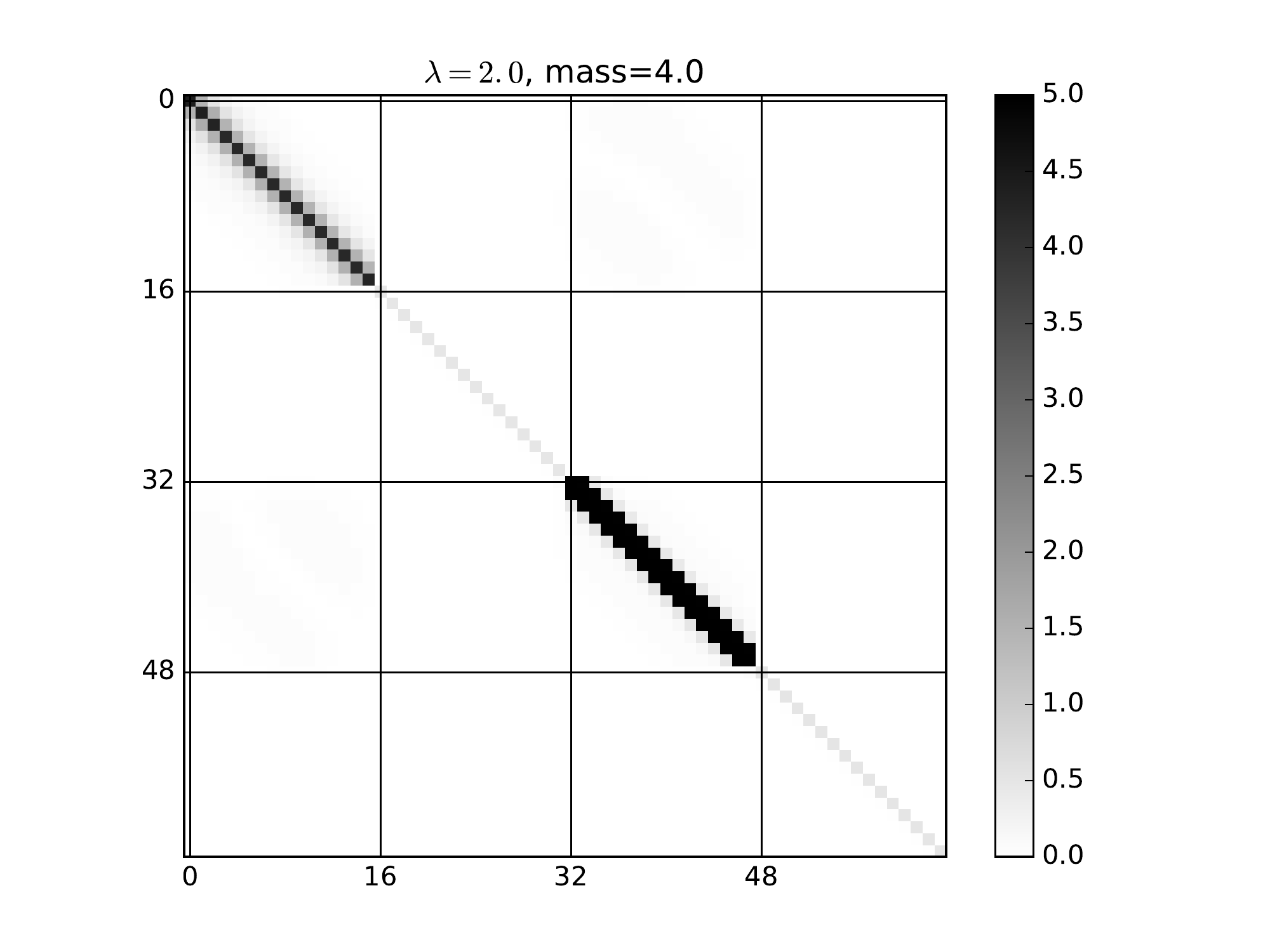}
\caption{Full matrix, $\lambda$=2.0, mass=4}
\label{fig:15}
\end{minipage}
\end{figure}

\section{summary, conclusions and outlook}

The purpose of this work is to examine the use of flow equation
methods to separate the physics on different resolution scales in an
exact wavelet discretization of quantum field theory.  While quantum
field theory couples all distance scales, there is a physically
relevant scale (or resolution) and it is desirable to formulate the
theory directly in terms of the degrees of freedom associated with the
physically relevant degrees of freedom.  This can be done by
eliminating the short-distance degrees of freedom.

While it may not be possible to get a well-defined local theory by
eliminating all arbitrarily small distance degrees of freedom, it is
possible formulate an effective theory that includes the important
short-distance physics by eliminating degrees of freedom between the
physically relevant scale and a chosen minimal resolution scale.  The
justification for this is that for a given application there is a
relevant volume and energy scale.  These restrictions generally imply
that the dynamics is dominated by a finite number of degrees of
freedom.  This can be understood in a number of ways.  For a free
field theory restricting the energy leads to a subspace of the Fock
space with an upper bound on the number of particles.  If this system
is put in a finite volume, the free particles in a finite volume 
have discrete energies and only a finite number of these states have 
energy less than the energy scale.  These degrees of freedom are 
sufficient to formulate the dynamics relevant to the system.

In the scaling-wavelet representation this provides a justification
for a volume/resolution truncation of the field theory.  
The smallest scales that influence the physics can be eliminated 
by block diagonalizing the Hamiltonian.  In this representation
the field was expressed as a linear combination of almost local 
operators classified by position and resolution.

These operators satisfy simple discrete canonical commutation
relations.  In this representation the exact Hamiltonian is a finite
degree polynomial with known constant coefficients in an infinite
number of these operators.  Resolution and volume truncations lead to
truncated Hamiltonians involving a finite number of degrees of
freedom.

The approach taken in this paper differs from other wavelet approaches
to quantum field theory\cite{federbush:1995}\cite{Battle:1999}
\cite{altaisky:2007}\cite{albeverio:2009}\cite{altaisky:2010}\cite{altaisky:2013}\cite{altaisky:2013b}\cite{altaisky:2016}.
The basis functions that are used in this work are orthonormal with
compact support, but they have a limited amount of smoothness.  In
most wavelet approaches the wavelet functions are over-complete, smooth
functions that do not have compact support.  The justification for
smearing fields with functions that have a limited amount of
smoothness is that when these basis functions are integrated against
free field Wightman functions the results are well-defined.  This
means the resulting operators are well defined on the free field Fock
space, and for theories truncated to a finite number of degrees of
freedom it is not necessary to pass to an in-equivalent representation
of the field algebra in order to solve the field equations.

While there are a number of potential methods to eliminate the short
distance degrees of freedom, the simple nature of the commutation
relations of the discrete field operators suggest using flow equation
methods, where the generator of the desired unitary transformation
involves commutators products of canonical pairs of operators.  These
methods have been proposed to be used in
QCD\cite{glazek93}\cite{glazek2}\cite{perryf} as well as in potential
theory \cite{bogner1}\cite{bogner2}\cite{perry}.  Those applications focus on momentum scales and
the equations are designed to drive the Hamiltonian to a diagonal
form.  In this work the goal is to formulate the problem in terms of
distance scales and to block diagonalize \cite{bartlett} 
the Hamiltonian rather than
diagonalize it.  In this example the transformed theory consists of
two sets of canonically conjugate set of operators that operate on
different distance scales.  The generator of the flow equation is
chosen to eliminate the terms that couple these scales in the
Hamiltonian.

This method was tested using the Hamiltonian for a free scalar
field.  While the free field is trivial, in the wavelet representation
the space derivatives in the Hamiltonian generate non-trivial terms
that couple the degrees of freedom on different scales.  This
Hamiltonian has the advantage that the flow equation applied to this
Hamiltonian does not generate an infinite number of new types of
many-body operators.  This allows us to focus on testing the flow
equation as a method to separate scales without the complication of
understanding the relative importance of the generated interactions.
What we found can be summarized by the following observations:

\begin{itemize}

\item[1.] It was demonstrated that flow equation methods with a suitable
  generator could be used to separate scales in a wavelet truncated
  theory.

\item[2.] The truncated free field Hamiltonian is equivalent to a
  system of coupled harmonic oscillators.  The flow equation block
  diagonalized the Hamiltonian with the coarse scale block containing
  the 16 lowest frequency normal modes and the fine scale block
  containing the 16 highest frequency modes.

\item[3.] Increasing the truncated volume generated new low frequency
  modes, while increasing the resolution increased the separation
  between modes.  The mass set a lower bound on the normal mode frequencies.

\item[4.] The flow equation exhibited convergence for masses between 0
  and 4.
  
\item[5.] For this problem, the flow equation was successfully applied
  directly to the Hamiltonian, without projecting on a subspace.

\item[6.] We found the that flow equation could be integrated using the
  Euler method, but perturbation theory failed to converge.

\item[7.] The evolved Hamiltonian was approximately local.   

\item[8.] The spectral properties suggest the in order to approach
the continuous spectrum of the exact theory, the volume and resolution
truncations need to be removed together.  

\item[9.] In our test the coefficients of the coupling terms initially
fell off quickly, but the rate of fall off slowed down significantly
as the flow parameter increased.  The method reduced the coupling
coefficients by a factors of about 100 for a modest value of the
flow parameter.

\end{itemize}  

The next step in this program is to consider models with interactions
and to consider models in 3+1 dimensions.  The complication with
interactions is that integrating the flow equation generates new
operators with each step of the Euler method.  A different flow
generator will be need to be formulated in order to get results
comparable to the results outlined above.  This is because the
analysis that led to (\ref{feq.20}-\ref{feq.21}) does not apply to the
interacting case.  On the other hand, the truncations suggest that in
the interacting case only a finite dimensional subspace of the Fock
space is relevant.  Flow equation methods are more naturally designed
to work on subspaces and diagonalize or block diagonalize operators
projected on subspaces.  These projections limit the types and nature
of the many-body operators that are generated by solving the flow
equations.  Generalizations to 3+1 dimensions are straightforward.
Single basis functions are replaced by products of three
basis functions.  While the bookkeeping becomes more difficult, the
basic structure is essentially unchanged.

The authors would like to thank Robert Perry and Mikhail Altaisky for
valuable feedback on this work.  We are also indebted to the referee,
who provided additional feedback and who pointed out the elegant
method form computing the overlap integrals due to Beylkin.  This work
was performed under the auspices of the U. S. Department of Energy,
Office of Nuclear Physics, award No. DE-SC0016457 with the University
of Iowa.

\section{appendix}

Daubechies wavelets and scaling functions are fractal functions.
Integrals involving these functions cannot be computed using
conventional methods, however the scaling equation (\ref{b.1}) and
normalization condition (\ref{b.2}) lead to linear constraints that
reduce the exact integration of all of the relevant integrals to
finite linear algebra.

These integrals were computed in \cite{fatih2}.  In this appendix we
calculate them using a method introduced by Beylkin \cite{beylkin-92}.
While the equations are identical to Beylkin's, we derive them without
computing second derivatives of the $K=3$ scaling functions, which
only have continuous first derivatives.

Repeated application of the scaling equation (\ref{b.1}) and the
definition of the mother wavelet (\ref{b.14}) can be used to express
the coefficients $D^k_{s;mn}, D^{kl}_{sw;mn}$ and $ D^{w;jl}_{mn}$ in
terms of $D^0_{s;mn}$ and the $h_l$ in table 1:
%\beq
\[
D^k_{s;mn} = \int s^{k\prime}_m(x) s^{k\prime}_n(x)
dx = 2^{2k} \int s' (x-m) s'(x-n) = 2^k D^0_{s;mn}
\]
%\label{a:1}
%\eeq
%\beq
\[
D^{kl}_{sw;mn} = \int s^{k\prime}_m(x) w^{l\prime}_n(x) dx =
2^{2(l+1)}\sum_{m'n'} H^{l+1-k}_{mm'}G_{nn'}D^0_{s;m'n'}
\]
%\label{a:2}
%\eeq
%\beq
\[
D^{w;jl}_{mn} = \int w^{j\prime}_m(x) w^{l\prime}_n(x) dx =
2^{2(l+1)}\sum_{m'n'} (GH^{l-j})_{mm'}G_{nn'} D^0_{s;m'n'} \qquad (l\geq j) 
\]
%\label{a:3} 
%\eeq
where the primes denote derivatives,  the matrices $H_{mn}$ and $G_{mn}$
are defined in terms of the
scaling function and wavelet weights by 
%\beq
\[
H_{mn} = h_{n-2m} \qquad G_{mn}= g_{n-2m}
\]
%\label{a:4}
%\eeq
and 
%\beq
\[
D^0_{s;mn} = \int s^{\prime}_m(x) s^{\prime}_n(x) dx .
\]
%\label{a:5}
%\eeq
Translational invariance 
implies that the $D^0_{s;mn}$ can be expressed in terms of  
%\beq
\[
D^0_{s;mn}= D^0_{s;0,n-m} .
\]
For $K=3$ there are nine non-zero coefficients 
\beq
D^0_{s;0m} := \int s'(x) s'(x-m) dx  \qquad -4 \leq m \leq 4.
\label{ap.1}
\eeq
Letting $x'= x-m$ gives $D^0_{s;0-m}=D^0_{s;0m}$ so there are only 
five independent integrals that need to be evaluated.  Differentiating the 
scaling equation (\ref{b.1}) gives a renormalization group  
equation for the derivatives 
\beq
s'(x-m) = 2 \sqrt{2} \sum_l h_l s'(2x-2m-l) .
\label{ap.2}
\eeq
Using (\ref{ap.2}) in (\ref{ap.1}) gives homogeneous equations relating the
non-zero $D^0_{s;0m}$s: 
\beq
D^0_{s;0m} = 4 \sum_{l,n} h_l h_{l+n} D^0_{s;02m+n}.
\label{ap.3}
\eeq
The coefficients 
\[
a_n: = 2 \sum_{l,n} h_l h_{l+n}
\]
are called autocorrelation coefficients.  It follows from the orthogonality
constraint on translates of the scaling function (\ref{b.4}) that $a_0=2.0$ and
$a_{2n}=0$ for $n\not=0$.  The remaining autocorrelation coefficients
for odd $n$ are rational.  For the Daubechies $K=3$ $h_l$ in table 1
the non-zero autocorrealtion coefficients can be computed and they are
\[
a_0=2 \qquad
a_1=a_{-1}={75\over 64} \qquad
a_3=a_{-3}=-{25\over 128} \qquad
a_5=a_{-5}={3\over 128}. 
\]
The homogeneous equation (\ref{ap.3}) can be expressed in terms of the $a_n$:
\beq
D^0_{s;0m} = \sum 2 a_n D^0_{s;02m+n} .
\label{ap.4}
\eeq
In order to solve this system for $D^0_{s;0m}$ an inhomogeneous
equation is needed.  To derive an inhomogeneous equation note that
for $K=3$,  1, $x$ and $x^2$ can be expressed pointwise as
locally finite expansions in the scaling functions.  The expansions 
have the form
\beq
1= \sum_n s_n(x)
\label{ap.5}
\eeq
\beq
x= \sum_n (n+<x>)  s_n(x)
\label{ap.6}
\eeq
\beq
x^2 = \sum_n (n^2 +2n<x> +<x^2>)s_n(x)
\label{ap.7}
\eeq
where $<x^n> = \int s(x)x^n$ are moments of the scaling function.
While the moments can also be computed exactly, they are not needed
to calculate the $D^0_{s;0m}$
Differentiating (\ref{ap.6}),  using (\ref{ap.5}), gives  
\[
1=\sum_n n s_{n}'(x)
\]
Differentiating (\ref{ap.7}) using (\ref{ap.6}) gives
\[
2x=
\sum_n (n^2 +2n<x>)s'_n(x) =
\sum_n n^2 s'_n(x) + 2<x>
\]
Multiply by $s'(x)$ and integrating gives
\[
\int 2xs'(x)= -2 = \sum_n n^2 \int s'_n(x)s'(x)dx + 2<x>\int s'(x) dx =
\sum_n n^2 \int s'_n(x)s'(x)dx .
\]
This gives the inhomogeneous equation
\beq
\sum_n n^2 D^0_{s;n0} = -2
\label{ap.8}
\eeq
The linear system consisting of equations (\ref{ap.4}) and (\ref{ap.8})
has rational coefficients and can be solved exactly for rational solutions
%\beq
%\[
%D^0_{s;mn} = D^0_{s;0,n-m}
%\]
%\label{a:14}
%\eeq
%where
%\beq
\[
D^0_{s;40}=D^0_{s;-40}=- 3/560 
\]
%\label{a:15}
%\eeq
%\beq
\[
D^0_{s;30}=D^0_{s;-30}= - 4/35 
\]
%\label{a:16}
%\eeq
%\beq
\[
D^0_{s;20}= D^0_{s;-20} =  92/105 
\]
%\label{a:17}
%\eeq
%\beq
\[
D^0_{s;10}= D^0_{s;-10} = -356/105 
\]
%\label{a:18}
%\eeq
%\beq
\[
D^0_{s;00} = 295/56 .
\]

Similar methods \cite{wavelets} can be used to compute the overlap
integrals (\ref{f.22}) that appear in interacting theories.

\end{document}